\documentclass[10pt,final,journal]{IEEEtran}

\usepackage{tipa}
\usepackage{amsfonts}
\usepackage{amsfonts}
\usepackage{amsfonts}
\usepackage{mathbbold}
\usepackage{amsfonts}
\usepackage{amssymb}
\usepackage{stfloats}
\usepackage{cite}
\usepackage{graphicx}
\usepackage{psfrag}
\usepackage{subfigure}
\usepackage{amsmath}
\usepackage{array}
\usepackage{multirow}

\DeclareMathOperator*{\argmin}{\arg\min}

\begin{document}

\title{Robust Lattice Alignment for $K$-user MIMO Interference Channels with Imperfect Channel Knowledge}

\newtheorem{Thm}{Theorem}
\newtheorem{Lem}{Lemma}
\newtheorem{Cor}{Corollary}
\newtheorem{Def}{Definition}
\newtheorem{Exam}{Example}
\newtheorem{Alg}{Algorithm}
\newtheorem{Prob}{Problem}
\newtheorem{Rem}{Remark}
\newtheorem{Proof}{Proof}
\newtheorem{Subproblem}{Subproblem}
\newtheorem{assumption}{Assumption}


\author{$^{\dag}$Huang~Huang,~\IEEEmembership{Student Member,~IEEE,
} $^{\dag}$Vincent K. N. Lau,~\IEEEmembership{Senior Member,~IEEE,}\\
$^{\ast}$Yinggang Du and $^{\ast}$Sheng Liu\\
$^{\dag}$ECE Department, Hong
Kong University of Science and Technology, Hong Kong \\
$^{\ast}$Huawei Technologies, Co. Ltd. China }


\maketitle

\begin{abstract}
In this paper, we consider a robust lattice alignment design for
$K$-user quasi-static MIMO interference channels with imperfect
channel knowledge. With random Gaussian inputs, the conventional
interference alignment (IA) method has the feasibility problem when
the channel is quasi-static. On the other hand, structured
lattices can create structured interference as opposed to the random
interference caused by random Gaussian symbols. The structured
interference space can be exploited to transmit the desired signals
over the gaps. However, the existing alignment methods on the
lattice codes for quasi-static channels either require {\em
infinite} SNR or {\em symmetric interference channel coefficients}.
Furthermore, perfect channel state information (CSI) is required for
these alignment methods, which is difficult to achieve in
practice. In this paper, we propose a robust lattice alignment
method for quasi-static MIMO interference channels with imperfect
CSI at all SNR regimes, and a two-stage decoding algorithm to decode
the desired signal from the {\em structured interference space}. We
derive the achievable data rate based on the proposed robust lattice
alignment method, where the design of the precoders,
decorrelators, scaling coefficients and {\em interference
quantization coefficients} is jointly formulated as a mixed
integer and continuous optimization problem. The effect of imperfect
CSI is also accommodated in the optimization formulation, and hence
the derived solution is robust to imperfect CSI. We also
design a low complex iterative optimization algorithm for our robust
lattice alignment method by using the existing iterative IA
algorithm that was designed for the conventional IA method.
Numerical results verify the advantages of the proposed robust
lattice alignment method compared with the TDMA, two-stage ML
decoding, generalized Han-Kobayashi (HK), distributive IA and
conventional IA methods in the literature.
\end{abstract}


\section{Introduction} Interference is a fundamental bottleneck in
wireless systems. This is partially due to the lack of understanding
interference from an information theoretic view. For example, the
capacity region of the two-user Gaussian interference channels has
been an open problem for over 30 years. It was only shown recently
that the Han-Kobayashi (HK) achievable region \cite{HK:gaussian_IC}
achieves the capacity region to within one
bit\cite{IC:gaussian:capacity}. Lately, there have been some
breakthroughs on the understanding of $K$-user interference
channels. In \cite{IA:conventional:2008} and
\cite{IA:M*N_MIMO:2008}, the authors propose an {\em interference
alignment} (IA) method to align interference onto a lower
dimensional subspace of each receiver so that the desired signal can
be transmitted on the {\em interference-free dimensions}. The
authors show that IA is optimal in the degrees-of-freedom (DoF)
sense (which is a high SNR performance measure) and the total
capacity of the $K$-user interference channels is given by
$\frac{K}{2}\log(\text{SNR})+o(\log(\text{SNR}))$
\cite{IA:conventional:2008}. There are a number of extensions
\cite{IA:distributed:2008,IA:alternative:2008} that have studied the
application of IA in $K$-user quasi-static MIMO interference
channels. However, the conventional IA method for $K$-user
interference channels requires infinite dimension in time-varying or
frequency-selective channels, and has the feasibility
problem. For example, it is shown in \cite{Feasibility:MIMO:2009}
that conventional IA in quasi-static MIMO ($M$ transmit and $N$
receive antennas) interference channels is not able to achieve a per
user DoF greater than $\frac{M+N}{K+1}$. As a result, there is no
satisfactory solution for quasi-static MIMO interference channels
for large $K$ due to the feasibility problem.

Besides the feasibility problem, conventional IA solutions have
assumed Gaussian input symbols and it is unclear whether it is
optimal to employ Gaussian input for interference channels. With
the Gaussian input symbols, the interference space is random and all
the conventional IA methods try to align all interference to a
smaller dimension space and utilize the remaining {\em
interference-free} dimensions to transmit the desired signals. While
Gaussian inputs (using random codebook argument) are capacity
optimal in multi-access and broadcast
channels\cite{Gaussian:BC:2006,Cover:06}, they are not optimal for
interference channels \cite{Many2one:2009,IC:gaussian:capacity}. It
is revealed in \cite{Many2one:2009} that structured interference is
more preferred in interference networks. Instead of giving up the
space that is reserved for the interference and transmitting the
desired signal on the remaining {\em interference-free dimensions},
one could also exploit the structured interference space and
transmit desired signals over the {\em gaps}. This motivates the
{\em lattices} based study of interference alignment. In
\cite{Symmetric:2008,Threeusers:2008}, the authors propose a lattice
interference alignment method to align the interfering lattices on a
common basis but it only works for symmetric SISO interference
channels (where all cross links have the same fading coefficients)
or a specialized class of 3-user SISO interference channels (where
the products of the fading coefficients are assumed to be rational).
In \cite{Real:2009,Real:MIMO:2009}, the authors propose interference
alignment over real line for $K$-user quasi-static MIMO
interference channels with real channel coefficients and demonstrate
that the DoF $=\frac{MN}{M+N}K$ can be achieved. This approach is
further extended to the complex channel in \cite{Complex:2009} for
the compound MIMO broadcast channel. In addition to interference
channels, lattice codes have also been widely studied in many
communication networks \cite{lattice:everywhere} such as
point-to-point channels and wireless relay networks. Specifically,
in \cite{lattice:AWGN:2004}, the author shows that lattice
codes can achieve the capacity of $\frac{1}{2}\log(1 + \text{SNR})$
in the point-to-point additive white Gaussian noise (AWGN) channel.
In \cite{CF:2009}, the authors exploit the lattice codes and propose
a compute-and-forward method in wireless relay networks. They show
that each relay can decode a linear function of messages from
multiple source nodes by using lattice codes. A destination node can
decode the desired message, given sufficient linear combinations
forwarded from the relay nodes.

All these results indicate that we can potentially benefit
from the {\em structured interference} created by lattice codes.
However, in order to have a practical method for exploiting the
advantage of the structured interference due to lattice
transmissions, there are still some key technical challenges to be
addressed.
\begin{itemize}
\item{\bf How to align the received lattices on a common basis over irrational quasi-static MIMO interference channels:}
Creating an ideal structured interference space
\cite{Symmetric:2008,Threeusers:2008} requires that all the
interfering lattices to be received on a common basis at each of the
$K$ receivers. However, this is a difficult requirement and is
impossible for irrational fading matrices. While
\cite{Symmetric:2008,Threeusers:2008} study lattice alignment
methods for $K$ user quasi-static interference channels, the methods
therein only work for {\em symmetric interference channels} or {\em
a specialized class of 3-user SISO interference channels}. These
approaches cannot be used for general non-symmetric irrational
interference channels.

\item {\bf How to exploit structured interference at finite SNR:}
In \cite{Real:2009,Real:MIMO:2009}, the authors propose a lattice
alignment algorithm for MIMO interference channels, which is optimal
in the DoF sense. However, it is not known whether this approach can
be applied at finite SNR, which is an important operating
regime in practice. In \cite{Ergodic:alignment:2009}, the
authors propose an ergodic interference alignment method which could
work at finite SNR regime. However, it requires either infinite time
or frequency extension and cannot be applied for constant channels.

\item {\bf How to ensure robustness due to imperfect knowledge of CSI:}
All of the above works assume perfect knowledge of channel state
information (CSI) for facilitating interference alignment. In
practice, this is not possible and interference alignment
performs very poorly with imperfect CSI \cite{imperfect:CSI:2009}.
It is quite challenging to incorporate robustness with respect to
(w.r.t.) imperfect CSI and exploit the structured interference space
at the same time.
\end{itemize}

In this paper, we explore a robust precoder and decorrelator design
for $K$-user general irrational quasi-static MIMO interference
channels with imperfect CSI. We propose a {\em robust lattice
alignment} method (which does not have the feasibility
problem) for general irrational constant interference channels, and
deduce a {\em two-stage decoding algorithm} to decode the desired
signal from the {\em structured interference space}. We derive the
achievable data rate based on the proposed method and formulate the
precoders, decorrelators, scaling coefficients as well as
the interference quantization coefficients\footnote{Please
refer to Section \ref{sec:problem} for details about these design
parameters.} design as a mixed integer and continuous optimization
problem \cite{liduan:2006}. By utilizing the alternating
optimization technique and by exploiting the structured
interference, we derive a low complexity algorithm to determine the
precoders, decorrelators as well as the interference quantization
coefficients. The effect of imperfect CSI is also accommodated in
the optimization formulation and hence, the derived solution is
robust to imperfect CSI. To illustrate the benefit of the proposed
method, we compare the achievable data rates with those of the
conventional IA method \cite{IA:conventional:2008}, the distributed
IA method based on alternating optimization
\cite{IA:distributed:2008}, the TDMA method, the brute-force
two-stage maximum likelihood (ML) decoding method and the
generalized HK method \cite{HK:gaussian_IC,IC:gaussian:capacity}
with random Gaussian input symbols. We show that the proposed method
achieves the same DoF as the conventional IA method if the problem
is feasible. On the other hand, when the conventional IA is not
feasible, our proposed solution can still offer significant
performance gain compared with these baselines.

This paper is organized as follows. In Section \ref{sec:model}, we
outline the system model of the $K$-user quasi-static MIMO
interference channels. In Section \ref{sec:preliminary}, we review
some common interference alignment methods in the literature and
provide some preliminary discussions on the lattices. In Section
\ref{sec:problem}, we propose the robust lattice alignment method
and formulate it as a mixed integer and continuous optimization
problem. In Section \ref{sec:alg}, by using the alternating
optimization technique, we derive a low complexity solution. In
Section \ref{sec:interpretation}, we derive closed-form analysis of
the proposed method and the baseline methods under specialized
channel realizations. The performance of the proposed method and the
existing methods under complex channel realizations is illustrated
in Section \ref{sec:sim}. We conclude with a brief summary of the
results in Section \ref{sec:con}.

\section{System Model}\label{sec:model}
\subsection{$K$-user Quasi-Static MIMO Interference Channels}
\begin{figure}
\centering
\includegraphics[width = 8.5cm]{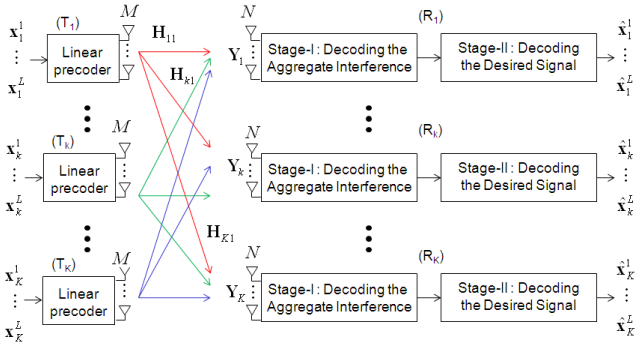}
\caption{Quasi-static complex $K$-user MIMO Gaussian interference
channels. Each transmitter, equipped with $M$ antennas, tries to
transmit $L$ independent data streams to its corresponding receiver,
which is equipped with $N$ antennas. } \label{fig:system_model}
\end{figure}

We consider the $K$-user quasi-static MIMO Gaussian interference
channels as illustrated in Fig. \ref{fig:system_model}. Each
transmitter, equipped with $M$ antennas, tries to communicate to its
corresponding receiver, which is equipped with $N$ antennas.
Specifically, we consider a block of $T$ channel symbols. The
channel output at the $k$-th receiver is described as follows:
\begin{equation}
\mathbf{Y}_k=\sum_{i}\mathbf{H}_{ki}\mathbf{\overline{X}}_i+\mathbf{Z}_k,
\end{equation}
where, $\mathbf{H}_{ki}$ is the $N\times M$ MIMO complex fading
coefficients from the $i$-th transmitter to the $k$-th receiver.
$\mathbf{\overline{X}}_i$ is the $M\times T$ complex signal vector
transmitted by transmitter $i$. $\mathbf{Z}_k$ is the $N\times T$
circularly symmetric AWGN vector at receiver $k$. We assume all
noise terms are i.i.d. zero mean complex Gaussian with unit
variance.

\subsection{Precoding at Transmitters}
In this paper, we assume that each transmitter transmits $L$
independent data streams. Specifically, $\mathbf{v}_i^l$ is the
$M\times1$ precoder for the $l$-th complex data stream
$\mathbf{x}_i^l$ ($1\times T$) at transmitter $i$. Therefore,
$\mathbf{\overline{X}}_i=\sum_{l}\mathbf{v}_i^l\mathbf{x}_i^l$. The
average power for each data stream is given by
$\frac{1}{T}\mathbb{E}[||\mathbf{x}_i^l||^2]=P$. The precoders are
chosen such that the total transmit power for each transmitter is no
more than $\gamma P$, i.e.,
\begin{equation}
\sum_{l=1}^L||\mathbf{v}_i^l||^2\leq \gamma, \quad \forall i,
\end{equation}
where
$||\mathbf{v}_i^l||^2=||\mathbf{v}_i^l||_2^2=(\mathbf{v}_i^l)^H\mathbf{v}_i^l$.

\subsection{Imperfect Channel State Information Model}
The imperfect CSI model can be expressed as:
\begin{equation}
\mathbf{\hat{H}}_{ki} = \mathbf{H}_{ki} +\triangle_{ki},
\end{equation}
where $\mathbf{\hat{H}}_{ki}$ is the estimated CSI that is known at
the central coordinator. $\triangle_{ki}$ is the CSI error
satisfying:
\begin{eqnarray}
\label{eq:csi_model}
\triangle_{ki}\in\mathcal{E}&=&\left\{\triangle_{ki}:
\parallel\triangle_{ki}\parallel_{F}^2\leq \epsilon^2 \right\}\nonumber\\&=&
\left\{\triangle_{ki}:\textrm{Tr}
\{\triangle_{ki}(\triangle_{ki})^H\}\leq \epsilon^2 \right\}.
\end{eqnarray}
This deterministic channel estimation error model is widely used in
the literature \cite{Robust:2009,Robust:2006,Robust:error:2006}.
Therefore, given the estimated CSI $\mathbf{\hat{H}}_{ki}$, the
uncertainty of $\mathbf{H}_{ki}$ is an ellipsoid centered at
$\mathbf{\hat{H}}_{ki}$ with radius $\epsilon$.

\section{ Summary of the Existing Interference Alignment Methods and
Lattices}\label{sec:preliminary}
\subsection{Interference Alignment on a Quasi-Static MIMO
Space}\label{sec:sub_IA} It is shown in \cite{IA:conventional:2008}
that in the 3-user quasi-static MIMO interference channels with
$M>1$ antennas, $\frac{3M}{2}$ DoF can be achieved. Specifically,
when $M$ is even, each transmitter sends $\frac{M}{2}$ independent
streams over the $M\times\frac{M}{2}$ precoder $\mathbf{V}_i$ for
transmitter $i$, i.e.,
\begin{equation}
\mathbf{\overline{X}}_i=\sum_{m=1}^{M/2}\mathbf{v}_{i}^m\mathbf{x}_{i}^m=\mathbf{V}_i\mathbf{X}_i.
\end{equation}
The precoders $\{\mathbf{V}_i\}_{i=1}^3$ are designed in terms of
the channel matrices, where the dimension of the interference space
is equal to $M/2$ at each receiver. Each receiver can simply cancel
the interference by zero-forcing and then decode the desired $M/2$
streams. When $M$ is odd, similar results can be obtained by
considering a two time-slot symbol extension of the channel, with
the same channel coefficients over the two symbols. However, it is
shown in \cite{Feasibility:MIMO:2009} that the above method is not
able to achieve a per user DoF greater than $\frac{M+N}{K+1}$.

\subsection{Interference Alignment on a Lattice}
In \cite{Threeusers:2008}, the authors consider IA on a lattice for
some special 3-user SISO interference channels $(M=N=1)$.
Specifically, the {\em real} channel fading coefficients have to
satisfy the following requirement:
\begin{equation}
\label{eq:csi_IA}
\frac{H_{12}}{H_{21}}\times\frac{H_{23}}{H_{32}}\times\frac{H_{31}}{H_{13}}=\frac{p}{q},
\end{equation}
where $p$ and $q$ are integers such that $gcd(p,q)=1$. $gcd(\cdot)$
is the greatest common divider of the integers. Transmitter $i$
chooses the lattice $\Lambda_i$ generated by using {\em construction
A} described in \cite{lattice:1997}. In order to align the
interfering lattices at each receiver, the lattices are chosen by:
\begin{equation}
\label{eq:review:lattice} H_{12}\Lambda_2=pH_{13}\Lambda_3,
H_{21}\Lambda_1=qH_{23}\Lambda_3,
H_{31}\Lambda_1=H_{32}\Lambda_2.
\end{equation}
As a result, a two-stage decoding algorithm can be done at each
receiver, which decodes the {\em aggregate interference} first and
then subtracts the aggregate interference from the received signal
to decode the desired information. Note that the symmetric channel
considered in \cite{Symmetric:2008} is a special case in
(\ref{eq:csi_IA}). This method requires strong interference channels
to decode the interference first (treating the desired signal as
additive noise). Furthermore, the requirement of adding the
interfering lattice on a common basis at each receiver as in
(\ref{eq:review:lattice}) becomes infeasible for general irrational
$K$-user quasi-static interference channels. A simple counter
example is given below.
\begin{Exam}[A Simple
Counter Example] \label{Exam:channel} A simple counter
example is given by: $
\mathbf{y}_1=\mathbf{x}_1+\sqrt{2}\mathbf{x}_2+\sqrt{3}\mathbf{x}_3+\mathbf{z}_1,
\mathbf{y}_2=\sqrt{5}\mathbf{x}_1+\mathbf{x}_2+\sqrt{7}\mathbf{x}_3+\mathbf{z}_2,
\mathbf{y}_3=\sqrt{11}\mathbf{x}_1+\sqrt{13}\mathbf{x}_2+\mathbf{x}_3+\mathbf{z}_3$.
It is not possible to align the interfering lattices on a common
basis at each of the 3 receivers. ~\hfill\IEEEQED
\end{Exam}

\subsection{Interference Alignment over the Real Line}
It is shown in \cite{Real:MIMO:2009} that $\frac{MN}{M+N}K$ DoF can
be achieved for quasi-static MIMO using IA on the {\em real} line
\cite{Real:2009}. This approach is extended to the complex channel
in \cite{Complex:2009} for the compound MIMO broadcast channel. Here
we review the basic idea for the SISO interference channels in
\cite{Real:2009}, and it is easy to extend to the MIMO case as in
\cite{Real:MIMO:2009,Complex:2009}. Specifically, the received
signal at receiver $k$ can be represented as \cite{Real:2009}
\begin{equation}
y_k=A\Biggl(\sum_{l=0}^{L_k-1}H_{kk}T_{kl}x_k^l+
\underbrace{\sum_{i=1,i\neq
k}^{K}\sum\nolimits_{l=0}^{L_i-1}H_{ki}T_{il}x_i^l}_{I_k}\Biggr)+z_k,
\end{equation}
where $A$ controls the input power of all users.
$x_k^l\in(-Q,Q)_{\mathbb{Z}}$ is one of the integers in the set
$(-Q,Q)$, and carries information for the $l$-th data stream for
user $k$. $T_{kl}$ is a constant real number, which is the direction
of the transmitted $l$-th data stream. It is chosen as monomials
with variables from channel coefficients. $I_k$ is the aggregated
interference caused by all users. The data streams are aligned if
they arrive at the same direction. e.g., interference $x_1^l$ and
$x_2^l$ are aligned at receiver 3 if $H_{31}T_1^l=H_{32}T_2^l$. By
carefully designing the transmit directions, it is shown in
\cite{Real:2009,Real:MIMO:2009} that $\frac{MN}{M+N}K$ DoF can be
achieved. However, this method requires infinite SNR and it is not
known if this method could be modified for finite SNR regime.

\subsection{Lattices}
In this section, we shall review some preliminaries on lattices from
\cite{CF:2009,lattice:everywhere} and the references therein. A
$T$-dimensional {\em lattice} $\Lambda$ is a set of points in
$\mathbb{R}^T$, and is given in terms of the lattice generator
matrix $\mathbf{L}\in\mathbb{R}^{T\times T}$:
\begin{equation}
\Lambda=\{\mathbf{x}=\mathbf{Lw}:\mathbf{w}\in\mathbb{Z}^T\}.
\end{equation}

A {\em lattice quantization} is a function,
$Q_{\Lambda}:\mathbb{R}^T\rightarrow\Lambda$, that maps a point
$\mathbf{x}$ to the nearest lattice point in Euclidean distance:
\begin{equation}
Q_{\Lambda}(\mathbf{x})=\arg\min_{\lambda\in\Lambda}||\mathbf{x}-\lambda||.
\end{equation}

The {\em fundamental Voronoi region}, $\mathcal{V}$, of $\Lambda$,
is the set of points in $\mathbb{R}^T$ closest to the zero vector,
i.e., $\mathcal{V}=\{\mathbf{x}:Q_{\Lambda}(\mathbf{x})=0\}$. The
modulo-$\Lambda$ operation w.r.t. the lattice is defined as
\begin{equation}
\mathbf{x}\text{ Mod } \Lambda = \mathbf{x}-Q_{\Lambda}(\mathbf{x}),
\end{equation}
which is also the quantization error of $\mathbf{x}$ w.r.t.
$\Lambda$.

A {\em nested lattice code} $\mathcal{L}_1$ is the set of all points
of a fine lattice $\Lambda_1$ that are within the fundamental
Voronoi region $\mathcal{V}$ of a {\em coarse lattice} $\Lambda$:
\begin{equation}
\mathcal{L}_1=\Lambda_1\cap\mathcal{V}=\{\mathbf{x}:\mathbf{x}=\lambda\text{
Mod }\Lambda,\lambda\in\Lambda_1\},
\end{equation}
where $\Lambda$ is said to be {\em nested} in $\Lambda_1$, i.e.,
$\Lambda\subseteq\Lambda_1$. Please refer to Fig. 2 in
\cite{CF:2009} for an illustration of nested lattice. The rate of a
nested lattice code is given by:
\begin{equation}
\label{eq:lattice_rate}
R=\frac{1}{T}\log(|\mathcal{L}_1|)=\frac{1}{T}\log\frac{\text{Vol}(\mathcal{V})}{\text{Vol}(\mathcal{V}_1)},
\end{equation}
where $\text{Vol}(\mathcal{V})$ is the volume of $\mathcal{V}$.

\section{Robust Lattice Alignment} \label{sec:problem} In this section, we
introduce the framework of our robust lattice alignment method and
the associated two-stage decoding algorithm. We derive the minimum
achievable data rate and design the precoder, decorrelator and the
interference quantization coefficients via an optimization approach.

\subsection{Encoding at Transmitters}
In this paper, we adopt the lattice encoding method, which is
used both in point-to-point channels \cite{lattice:AWGN:2004} and
relay networks \cite{CF:2009}. Specifically, the data stream
$\mathbf{x}_k^l$ is given by:
\begin{equation}\label{eq:encoding}
\mathbf{x}_k^l=\left[\mathbf{t}_k^l-\mathbf{d}_k^l \right] \text{Mod
} \Lambda + j
\left[\mathbf{\widetilde{t}}_k^l-\mathbf{\widetilde{d}}_k^l \right]
\text{Mod } \Lambda,
\end{equation}
where $\Lambda\in\mathbb{R}^T$ is the coarse lattice used in the
nested lattice encoding method for the transmitted symbols
$\mathbf{x}_k^l, \forall k,l$. $\{\mathbf{t}_k^l,
\mathbf{\widetilde{t}}_k^l\}$ are points in the nested lattice code
$\mathcal{L}_k^l$ (corresponding to the fine lattice
$\Lambda_{k}^l$) that carry information.
$\{\mathbf{d}_k^l,\mathbf{\widetilde{d}}_k^l\}$ are the dither
vectors \cite{lattice:AWGN:2004,CF:2009} which are independently
uniformly distributed over $\mathcal{V}$ and are available to all
transmitters and receivers. Specifically, all the nested lattices
$\{\Lambda_{k}^l\}$ are AWGN good, and the coarse lattice $\Lambda$
is quantization good\footnote{Please refer to
\cite{lattice:everywhere} for the definition of goodness for lattice
codes. Specifically, the lattice codes should have both good
statistical and good algebraic properties\cite{CF:2009}.}. In
\cite{CF:2009}, such lattices and dither vectors satisfying the
transmit power constraint
$\frac{1}{T}\mathbb{E}[||\mathbf{x}_k^l||^2]=P$ are proved to exist.
\begin{Rem}[Complexity of the Encoding Method]
The complexity of the encoding method is similar to that in the
baselines. In the baseline methods, standard Gaussian random
codebook is assumed to achieve the mutual information rate. On
the other hand, there exists efficient coding schemes by using a
scalar constellation coupled with a linear code and this could come
close to the nested lattice performance (without the shaping
gain)\cite{Erez:05}. ~\hfill\IEEEQED
\end{Rem}

Note that, dithering is a common randomization technique in lattice
quantization for source coding \cite{lattice:AWGN:2004}. The
following two lemmas from \cite{lattice:AWGN:2004,CF:2009} capture
the key properties of the dithered nested lattice codes.
\begin{Lem}[Erez-Zamir\cite{lattice:AWGN:2004}]\label{lem:erez-zamir} For any random variable
$\mathbf{t}\in\mathcal{V}$, if $\mathbf{d}$ is statistically
independent of $\mathbf{t}$ and uniformly distributed over
$\mathcal{V}$, $[\mathbf{t}-\mathbf{d}]\text{ Mod } \Lambda$ is
uniformly distributed over $\mathcal{V}$, and is statistically
independent of $\mathbf{t}$. ~\hfill\IEEEQED
\end{Lem}
\begin{Lem}[Nazer-Gastpar\cite{CF:2009}]\label{lem:nazer-gastpar} Let $\mathbf{z}\sim \mathcal{N}(\mathbf{0},\mathbf{I}_{T\times
T})$ and $\mathbf{d}_i^n$ be statistically independently uniformly
distributed over $\mathcal{V}$ with
$P=\frac{1}{T}\mathbb{E}[||\mathbf{d}_i^n||^2]$, and
$\mathbf{z}_k^l=\alpha\mathbf{z}+\sum_{i,n}\theta_{i}^n\mathbf{d}_i^n$
for some constant $\alpha$ and $\theta_{i}^n$. The density of
$\mathbf{z}_k^l$ is upper bounded by the density of an i.i.d.
zero-mean Gaussian vector $\mathbf{\widetilde{z}}_k^l$ whose
variance approaches $N_k^l$ as $T\rightarrow\infty$, where $N_k^l =
\alpha^2+P\sum_{i,n}(\theta_{i}^n)^2$.      ~\hfill\IEEEQED
\end{Lem}
\begin{Rem}[Interpretation of the Lemmas]
Lemma \ref{lem:erez-zamir} assures that the input power exactly
meets the power constraint \cite{lattice:AWGN:2004}. Lemma
\ref{lem:nazer-gastpar} indicates the non-Gaussian noise
$\mathbf{z}_k^l$ is nearly Gaussian as the code length $T$
increases. ~\hfill \IEEEQED
\end{Rem}

\subsection{Lattice Alignment with Imperfect CSI}
\begin{figure}
\centering
\includegraphics[width = 8.5cm]{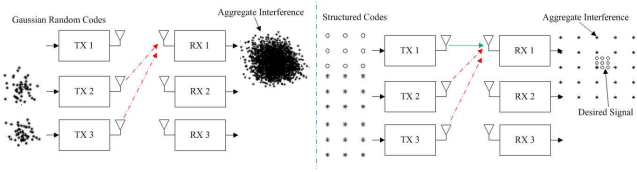}
\caption{Illustration of random interference (due to random
codes) and structured interference (due to structured codes) in
interference channels \cite[Fig. 4, Fig. 5]{Many2one:2009}. The
resulting random interference covers the entire space, preventing
the desired receiver from decoding. On the other hand, the
structured interference allows the desired receiver to decode
between gaps. }\label{fig:interference_structure}
\end{figure}

In this section, we shall discuss the proposed robust lattice
alignment method (using vector space strategies) and the
motivations. When Gaussian signals are transmitted in interference
channels, the associated interference space is random as illustrated
in Fig. \ref{fig:interference_structure}\cite{Many2one:2009}.
Intuitively, the aggregate interference will fill the entire signal
space and there is no room left for desired signal transmission.
Furthermore, the {\em penalty} of interference scales with the
number of interferers. As a result, the key idea behind the IA
methods in \cite{IA:conventional:2008,IA:M*N_MIMO:2008} is to align
all interferences to a lower dimensional subspace and utilize the
remaining interference free dimensions to transmit the desired
signals. However, one problem associated with the IA method in
\cite{IA:conventional:2008,IA:M*N_MIMO:2008} is the feasibility
problem. In fact, it is shown in \cite{Feasibility:MIMO:2009} that
the per user DoF greater than $\frac{M+N}{K+1}$ is not achievable.

On the other hand, when structured lattices are used at the
transmitters, the aggregate interference at each of the receiver may
be structured (if the interfering lattices are properly aligned) as
illustrated in Fig.
\ref{fig:interference_structure}\cite{Many2one:2009}. In this case,
the interference space contains regular {\em gaps} that can be
utilized to transmit desired signals. More importantly, the {\em
penalty} of the interference does not scale with the number of
interferers. Therefore, another possible direction of alignment is
to align the transmit lattices on a common basis at each of the $K$
receivers to create a structured interference space as illustrated
in Fig. \ref{fig:interference_structure}\cite{Many2one:2009}.
However, it is not always possible to simultaneously align all
the interfering lattices, as shown in Example \ref{Exam:channel}.
Furthermore, perfect lattice alignment requires perfect CSI, which
is impractical.

\begin{figure}
\centering
\includegraphics[width = 8.5cm]{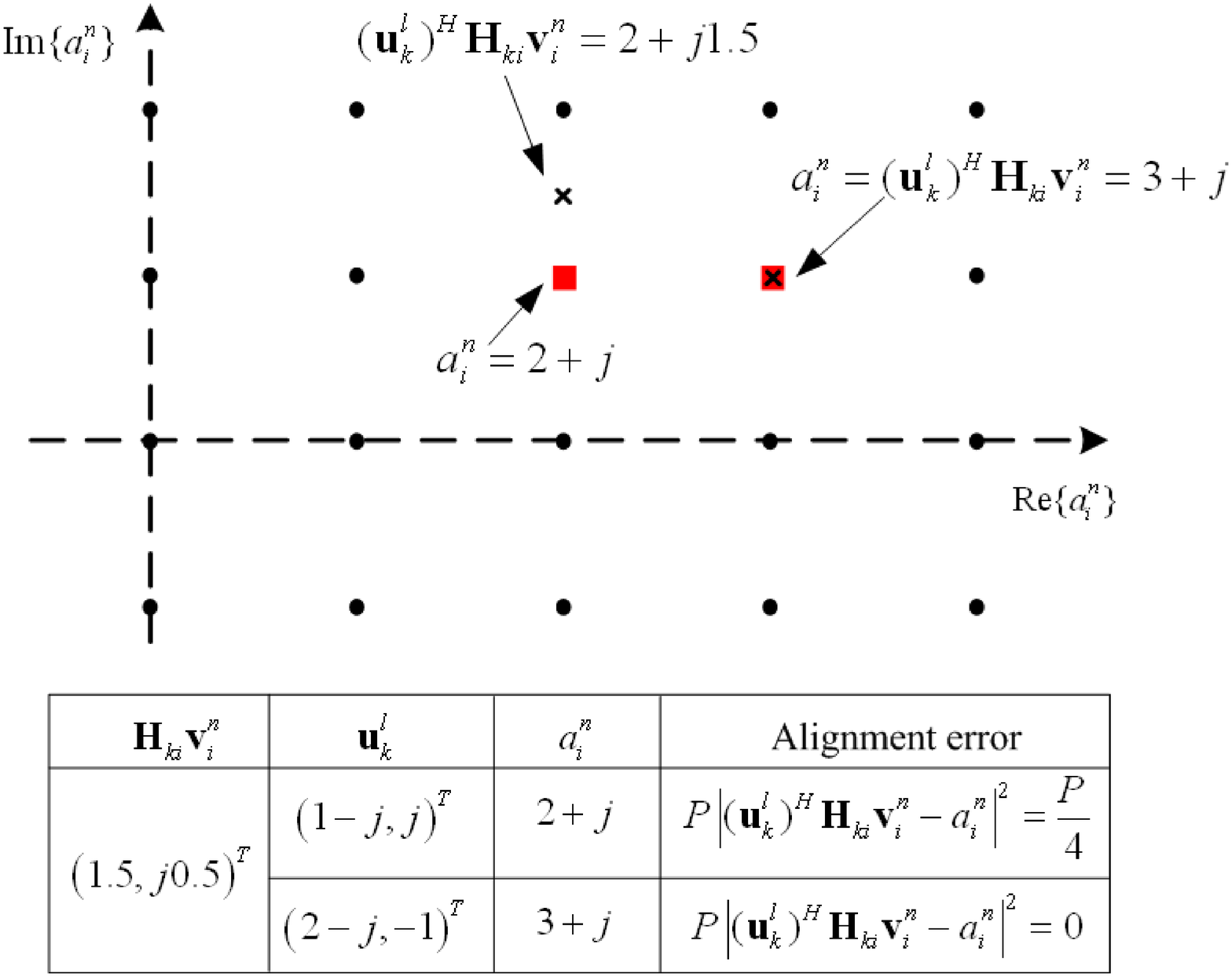}
\caption{Illustration of residual alignment error after stage I
decoding and how it is affected by different choices of stage I
decorrelator $\mathbf{u}_k^l$ and interference quantization
coefficients $\mathbf{a}=\{\{a_i^n\}_{i=1}^K\}_{n=1}^L$.
Specifically, $a_i^n$ is the interference quantization coefficient
and $\mathbf{H}_{ki}\mathbf{v}_i^n$ is the equivalent channel for
the stream $\mathbf{x}_i^n$ respectively.} \label{fig:align_error}
\end{figure}

Motivated by the advantages of having structured interference, we
propose a {\em robust lattice alignment method} in which the
precoders are designed to {\em align} the received lattices as much
as possible. We accept the fact that lattice alignment may not be
perfect (due to infeasible irrational channel coefficients and
imperfect CSI) and we try to minimize the effect of the {\em
residual lattice alignment error}. Specifically, instead of
decoding and subtracting each interference individually, we wish to
decode the structured aggregate interference composed of all the
interferences and remove them all at once. Furthermore, due to the
structural constraints of the lattice, we must choose integer
coefficient for each data stream to approximate the equivalent
channel coefficient. Therefore, the {\em aggregate interference
lattice} for the data stream $\mathbf{x}_k^l$ is written as:
\begin{equation}
\label{eq:lattice_interference} \mathbf{I}_k^l=
\sum\nolimits_{i,n}a_{i}^n\mathbf{x}_i^n,
\end{equation}
where $a_i^n\in\mathbb{Z}+j\mathbb{Z}$ (complex integer) is defined
as the {\em interference quantization coefficient} to approximate
the equivalent channel coefficient for $\mathbf{x}_i^n$, and
$a_k^l=0$ (see Fig. \ref{fig:align_error} for an illustration).
Specifically, given the precoder $\mathbf{v}_i^n$ for
$\mathbf{x}_i^n$, the decorrelator $\mathbf{u}_k^l$ at receiver $k$,
and the interference quantization coefficients $\{a_i^n\}$, the
alignment error is given by:
\begin{equation}
\label{eq:error_align} \mathbf{I}_e=P\sum\nolimits_{i\neq k\atop
n\neq
l}\Bigl|(\mathbf{u}_k^l)^H\mathbf{H}_{ki}\mathbf{v}_i^n-a_{i}^n\Bigr|^2.
\end{equation}

\begin{Rem}[The Effect of Design Parameters on $\mathbf{I}_e$]
Note that if the channel coefficients
$\{\mathbf{H}_{ki}\}_{i,k=1}^K$ are irrational, the alignment error
$\mathbf{I}_e$ may not be zero but the design parameters
$\{\mathbf{u}_k^l,\mathbf{v}_i^n,a_i^n\}$ in (\ref{eq:error_align})
can be chosen to minimize the effect of the alignment error as
illustrated in Fig. \ref{fig:align_error}. We shall elaborate the
precise optimization problem in Section \ref{sec:sub_problem}.
~\hfill\IEEEQED
\end{Rem}

\begin{figure}
\centering
\includegraphics[width = 8.6cm]{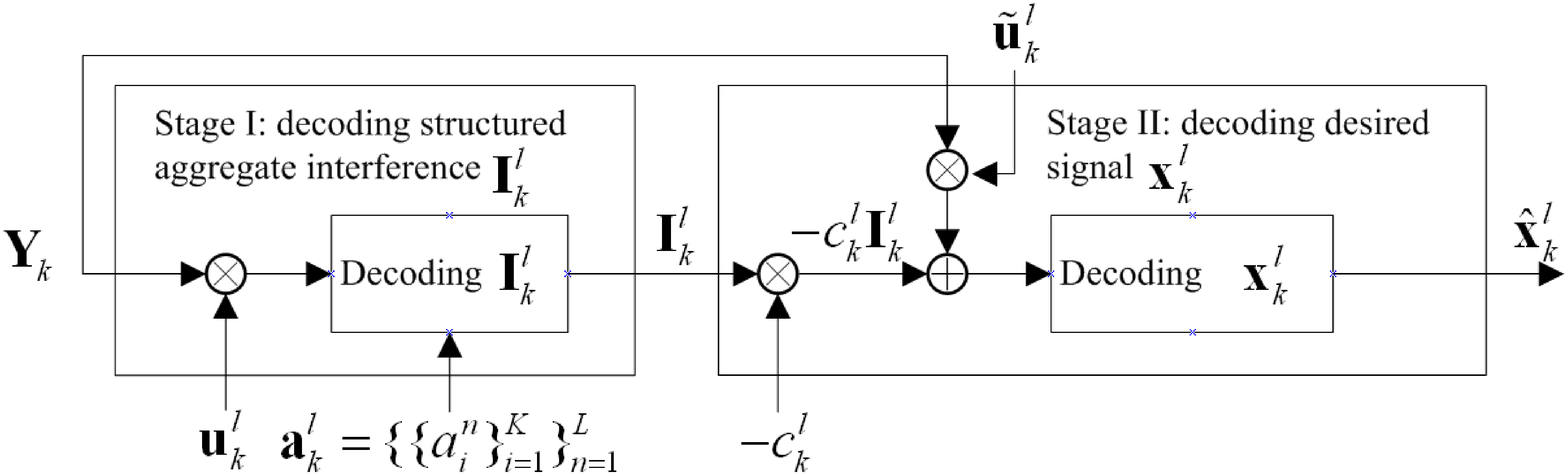}
\caption{The block diagram of the two-stage decoding algorithm for
decoding the desired signal $\mathbf{x}_k^l$. Specifically,
$\mathbf{Y}_k$ is the observed signal for user $k$ given by
(\ref{eq:y_k}), $\mathbf{I}_k^l$ is the structured aggregate
interference given by (\ref{eq:i_k^l}), $\mathbf{u}_k^l$ is the
first stage decorrelator, $\mathbf{a}_k^l$ are the interference
quantization coefficients, $c_k^l$ is the scaling coefficient, and
$\widetilde{\mathbf{u}}_k^l$ is the second stage decorrelator.}
\label{fig:decoding_algorithm}
\end{figure}

 Based on the {\em aggregate interference lattice} in
(\ref{eq:lattice_interference}), a {\em two-stage decoding
algorithm} can be constructed at each of the receivers as
illustrated in Fig. \ref{fig:decoding_algorithm}. The first stage
is to decode the structured aggregate interference, whereas the
second stage is to decode the desired signal after canceling the
decoded aggregate interference. The decoding process and the
associated error analysis in the presence of the residual alignment
errors are discussed in detail in the following subsections.

\subsection{Stage I - Decoding the Aggregate Interference}
\begin{figure*}[!t]
\normalsize
\begin{equation}
\label{eq:R_k1}
R_{i}^{n}<\mu_k^l=
\left\{\begin{array}{ll}
\log\Big(\frac{P}{||\mathbf{u}_k^l||^2+
P\sum\limits_{i,n}\big||(\mathbf{u}_k^l)^H\hat{\mathbf{H}}_{ki}\mathbf{v}_i^n-a_{i}^n|+ \epsilon||\mathbf{v}_i^n||\cdot||\mathbf{u}_k^l||\big|^2}\Big)
& \text{if } a_i^n\neq 0\\
\infty & \text{if } a_i^n= 0
\end{array}\right.,
\end{equation}
\begin{eqnarray}
\label{eq:R_k2}
R_{k}^{l}<\widetilde{\mu}_{k}^l=\log\Big(\frac{P}{||\widetilde{\mathbf{u}}_k^l||^2+
P\sum\nolimits_{i,n}\bigl||(\widetilde{\mathbf{u}}_k^l)^H\hat{\mathbf{H}}_{ki}\mathbf{v}_i^n-c_k^la_{i}^n-1_{\{i=k\&n=l\}}|+ \epsilon||\mathbf{v}_i^n||\cdot||\widetilde{\mathbf{u}}_k^l||\bigr|^2}\Big).
\end{eqnarray}
\hrulefill
\end{figure*}

Note that the received signal at the $k$-th receiver is given by:
\begin{equation}
\label{eq:y_k}
\mathbf{Y}_k=\mathbf{H}_{kk}\mathbf{v}_k^l\mathbf{x}_k^l+\sum_{n\neq
l}\mathbf{H}_{kk}\mathbf{v}_k^n\mathbf{x}_k^n+\sum_{i\neq
k}\sum_{n}\mathbf{H}_{ki}\mathbf{v}_i^n\mathbf{x}_i^n+\mathbf{Z}_k.
\end{equation}

At stage I, a complex linear decorrelator $\mathbf{u}_k^l$
($N\times1$) is applied at receiver $k$, i.e.,
\begin{equation}
\label{eq:y_k^l_1st}
\begin{array}{l}
\mathbf{y}_k^l=(\mathbf{u}_k^l)^H\mathbf{Y}_k=
\underbrace{(\mathbf{u}_k^l)^H\mathbf{H}_{kk}\mathbf{v}_k^l\mathbf{x}_k^l}_{\text{desired
signal}}+ \underbrace{\sum_{n\neq
l}(\mathbf{u}_k^l)^H\mathbf{H}_{kk}\mathbf{v}_k^n\mathbf{x}_k^n}_{\text{inter-stream
interference}} \\
\quad\quad+\underbrace{\sum_{i\neq
k;n}(\mathbf{u}_k^l)^H\mathbf{H}_{ki}\mathbf{v}_i^n\mathbf{x}_i^n}_{\text{inter-user
interference}} +(\mathbf{u}_k^l)^H\mathbf{Z}_k.
\end{array}
\end{equation}

From $\mathbf{y}_k^l$ ($1\times T$), we wish to decode the
structured aggregate interference, i.e.,
\begin{equation}
\label{eq:i_k^l}
\mathbf{I}_{k}^l=\left[\sum\nolimits_{i,n}\Re\{a_{i}^n\mathbf{x}_i^n\}\right]\text{mod
}\Lambda + j
\left[\sum\nolimits_{i,n}\Im\{a_{i}^n\mathbf{x}_i^n\}\right]
\text{mod }\Lambda,
\end{equation}
where $a_{i}^n\in\mathbb{Z}+j\mathbb{Z}$, and $a_{k}^l=0$. Let
$\mathbf{a}_k^l=\{\{a_{i}^n\}_{i=1}^K\}_{n=1}^L$ denote the {\em
interference quantization coefficients} for decoding the desired
data stream $\mathbf{x}_k^l$. $\Re\{\cdot\}$ and $\Im\{\cdot\}$
denote the real and imaginary parts, respectively. To successfully
decode $\mathbf{I}_{k}^l$, there are some requirements on the
transmit data rates of the streams in the $\mathbf{I}_{k}^l$ term.
The achievable rate region of the $K$ users for successful stage I
decoding under the imperfect CSI model in (\ref{eq:csi_model}) is
summarized in Lemma \ref{lem:rate}.
\begin{Lem}\label{lem:rate} {\em (A Sufficient Condition for Successful Stage I Decoding under Imperfect CSI):}
Under the imperfect CSI model in (\ref{eq:csi_model}), the
structured aggregate interference $\mathbf{I}_{k}^l$ can be decoded
from $\mathbf{y}_k^l$ with arbitrarily small error probability if
the data rate $R_i^n$ satisfies the condition (for all
$i,n$) given in \eqref{eq:R_k1}, where $||\mathbf{v}_i^n||\cdot||\mathbf{u}_k^l||$ means the product
of the two terms $\{||\mathbf{v}_i^n||,||\mathbf{u}_k^l||\}$.
~\hfill\IEEEQED
\end{Lem}
\begin{proof}
Please refer to Appendix \ref{app:rate}. Note that a similar result
with perfect CSI is derived in \cite{CF:2009} for single-stream
wireless relay networks.
\end{proof}

Note that when $a_i^n = 0$ for all $\{i,n\}$, i.e.,
$\mathbf{a}_k^l=\mathbf{0}$, there will be no stage I decoding for
the desired data stream $\mathbf{x}_k^l$ and the proposed method
reduces to the conventional precoder-decorrelator optimization with
one-stage decoding. The first term and the second term in the
denominator of (\ref{eq:R_k1}) correspond to the noise term and the
residual lattice alignment errors, respectively.

\subsection{Stage II - Decoding the Desired Signal}
After successfully decoding the structured aggregate interference
$\mathbf{I}_k^l$ at stage I, we wish to decode the desired signal
$\mathbf{x}_k^l$ at stage II. This is illustrated in Fig.
\ref{fig:decoding_algorithm}. Specifically, the scaled {\em
structured aggregate interference} $c_k^l\mathbf{I}_{k}^l$ is
subtracted from the received signal and the desired signal is
decoded via $\mathbf{\widetilde{y}}_k^l$ given by:
\begin{eqnarray}
\label{eq:2nd_y_k}
\mathbf{\widetilde{y}}_k^l=\left\{\begin{array}{ll}
(\widetilde{\mathbf{u}}_k^l)^H\mathbf{Y}_k-(\widetilde{\mathbf{u}}_k^l)^H\mathbf{H}_{km}\mathbf{v}_m^d\mathbf{x}_m^d&\text{If
} \mathbf{a}_k^l=\delta_{m}^d \\
\left[\Re\bigl\{(\widetilde{\mathbf{u}}_k^l)^H\mathbf{Y}_k-c_k^l\mathbf{I}_{k}^l\bigr\}\right]\text{Mod
}\Lambda
+\atop \quad j\left[\Im\bigl\{(\widetilde{\mathbf{u}}_k^l)^H\mathbf{Y}_k-c_k^l\mathbf{I}_{k}^l\bigr\}\right]\text{Mod
}\Lambda& \text{otherwise}\\
\end{array}\right.,
\end{eqnarray}
where $\widetilde{\mathbf{u}}_k^l$ is the second stage decorrelator,
$c_k^l\in\mathbb{Z}+j\mathbb{Z}$ is the {\em scaling
coefficient}\footnote{ We may need to scale the decoded aggregate
interference $\mathbf{I}_k^l$ first before subtracting from the
received signal so as to compensate for the amplitude change due to
the second stage decorrelator $\widetilde{\mathbf{u}}_k^l$.} for the
$l$-th data stream at receiver $k$, $\delta_{m}^d$ is the
$\delta$-function (all $a_i^n=0$ except $a_m^d=1$), and
\begin{equation}
\label{eq:2nd_y_k_xx_real}
\begin{array}{l}
\quad\left[\Re\bigl\{(\widetilde{\mathbf{u}}_k^l)^H\mathbf{Y}_k-c_k^l\mathbf{I}_{k}^l\bigr\}\right]\text{Mod
}\Lambda\\
=\Big[\Re\big\{(\widetilde{\mathbf{u}}_k^l)^H\mathbf{Y}_k\big\}-\Re\big\{c_k^l\big\}\Re\big\{\mathbf{I}_{k}^l\big\}
+\Im\big\{c_k^l\big\}\Im\big\{\mathbf{I}_{k}^l\big\}\Big]\text{Mod
}\Lambda \\
\stackrel{(a)}{=}\Big[\Re\big\{(\widetilde{\mathbf{u}}_k^l)^H\mathbf{Y}_k\big\}- \big[\Re\big\{c_k^l\big\}\Re\big\{\mathbf{I}_{k}^l\big\}\big]\text{Mod
}\Lambda\\
\quad\quad\quad +\big[\Im\big\{c_k^l\big\}\Im\big\{\mathbf{I}_{k}^l\big\}\big]\text{Mod
}\Lambda \Big]\text{Mod
}\Lambda\\
\stackrel{(b)}{=}\Big[\Re\big\{\sum\nolimits_{i,n}[(\widetilde{\mathbf{u}}_k^l)^H\mathbf{H}_{ki}\mathbf{v}_i^n-c_k^la_{i}^n]
\mathbf{x}_i^n+\widetilde{\mathbf{u}}_k^l\mathbf{Z}_k\big\}\Big]
\text{Mod }\Lambda
\end{array}
\end{equation}
where $(a)$ follows from $\big[ \mathbf{g}_1+\mathbf{g}_2
\big]\text{Mod }\Lambda = \big[ ( [\mathbf{g}_1]\text{Mod }\Lambda )
+\mathbf{g}_2 \big]\text{Mod }\Lambda$ for all
$\mathbf{g}_1,\mathbf{g}_2\in\mathbb{R}^T$, $(b)$ follows from
$c_k^l\in\mathbb{Z}+j\mathbb{Z}$ and $\big[ c(
[\mathbf{g}_1]\text{Mod }\Lambda )\big]\text{Mod }\Lambda=\big[
c\mathbf{g}_1\big]\text{Mod }\Lambda$ for all $c\in\mathbb{Z}$.
Similarly, we have
\begin{equation}
\begin{array}{l}\label{eq:2nd_y_k_xx_imag}
\left[\Im\bigl\{(\widetilde{\mathbf{u}}_k^l)^H\mathbf{Y}_k-c_k^l\mathbf{I}_{k}^l\bigr\}\right]\text{Mod
}\Lambda=\\
\Big[\Im\big\{\sum\nolimits_{i,n}[(\widetilde{\mathbf{u}}_k^l)^H\mathbf{H}_{ki}\mathbf{v}_i^n-c_k^la_{i}^n]
\mathbf{x}_i^n+\widetilde{\mathbf{u}}_k^l\mathbf{Z}_k\big\}\Big]
\text{Mod }\Lambda.
\end{array}
\end{equation}

From (\ref{eq:2nd_y_k}) note that, if we set
$\mathbf{a}_k^l=\delta_{m}^d$, then $\mathbf{I}_k^l=\mathbf{x}_m^d$,
and hence the stage II decoding in (\ref{eq:2nd_y_k}) would try to
directly null off the interference $\mathbf{x}_m^d$. Similarly, the
sufficient condition in terms of the rate region for successful
stage II decoding is given in the next lemma:
\begin{Lem} {\em (A Sufficient Condition for Successful Stage II Decoding under Imperfect CSI):}
\label{lem:rate2} A sufficient condition for successful
stage II decoding under the imperfect CSI model in
(\ref{eq:csi_model}) is given in \eqref{eq:R_k2},
where $1_{\{\cdot\}}$ is an indicator function. ~\hfill\IEEEQED
\end{Lem}
\begin{proof}
The proof follows from a similar approach as in Appendix
\ref{app:rate}. Specifically, it is obtained by replacing
$\mathbf{y}_k^l$ and $\mathbf{I}_k^l$ (obtained from
(\ref{eq:2nd_y_k_xx_real}) and (\ref{eq:2nd_y_k_xx_imag})) with
$\mathbf{\widetilde{y}}_k^l$ and $\mathbf{x}_k^l$ in the proof of
Lemma \ref{lem:rate}, respectively. Note that, from
(\ref{eq:2nd_y_k_xx_real}) and (\ref{eq:2nd_y_k_xx_imag}), the
equivalent channel coefficient for data stream $\mathbf{x}_i^n$ in
stage II decoding is
$(\widetilde{\mathbf{u}}_k^l)^H\mathbf{H}_{ki}\mathbf{v}_i^n-c_k^la_{i}^n$.
Furthermore, since we wish to decode
$\mathbf{x}_k^l=1\cdot\mathbf{x}_k^l+\sum_{i\neq k\atop n\neq
l}0\cdot\mathbf{x}_i^n$, $1_{\{i=k\& n=l\}}$ in (\ref{eq:R_k2})
indicates the coefficient for $\mathbf{x}_k^l$ and the coefficients
for the other data streams are all 0.
\end{proof}

Note that the rate region in (\ref{eq:R_k2}) is a function of
interference quantization coefficients $\mathbf{a}$, scaling
coefficients $\mathbf{c}=\{\{c_k^l\}_{k=1}^K\}_{l=1}^L$, precoders
$\mathbf{v}$, and second stage decorrelators
$\widetilde{\mathbf{u}}=\{\{\widetilde{\mathbf{u}}_k^l\}_{k=1}^K\}_{l=1}^L$.
\begin{Rem}[Complexity of the Decoding Method]
The decoding complexity of the proposed design and the baselines are
similar. For instance, all baseline methods have assumed ML decoding
to achieve the mutual information rate. On the other hand, there
exists efficient lattice decoding methods
\cite{lattice:AWGN:2004,CF:2009} that could exploit the lattice
symmetry in the decoding process. ~\hfill\IEEEQED
\end{Rem}

\subsection{Precoder, Decorrelator and Interference Quantization Coefficients Optimization}\label{sec:sub_problem}
In this subsection, we shall formulate the precoders, decorrelators,
scaling coefficients and interference quantization coefficients
design problem formally as an optimization problem. The problem
consists of the following components:
\begin{itemize}
\item{\bf Optimization Variables:} Specifically, the optimization
variables consist of the set of precoders $\mathbf{v}$, the set of
stage I decorrelators $\mathbf{u}$, the set of stage II
decorrelators $\widetilde{\mathbf{u}}$, the set of interference
quantization coefficients $\mathbf{a}$ (used in stage I processing),
and the set of scaling coefficients $\mathbf{c}$  (used in stage II
processing).

\item{\bf Optimization Objective:} For fairness, we
consider the worst-case data rate as the optimization objective,
i.e.,
$\max\limits_{\mathbf{u},\widetilde{\mathbf{u}},\mathbf{v},\mathbf{a},\mathbf{c}}R_{\min}
$, where $R_{\min} =
\min_{k,l}\left(\mu_k^l,\widetilde{\mu}_k^l\right)$.

\item{\bf Optimization Constraints:} The optimization constraints
consist of the transmit power constraint
$\sum_{l=1}^L||\mathbf{v}_k^l||^2\leq \gamma, \forall k$, the
interference quantization coefficients constraint
$\mathbf{a}_{k}^l\in(\mathbb{Z}+j\mathbb{Z})^{KL}$, and the scaling
coefficients constraint $c_{k}^l\in\mathbb{Z}+j\mathbb{Z}$.

\end{itemize}

As a result, the optimization problem is summarized below:
\begin{equation}
\label{eq:maximin}
\{\mathbf{u}^*,\widetilde{\mathbf{u}}^*,\mathbf{v}^*,\mathbf{a}^*,\mathbf{c}^*\}=\left\{\begin{array}{l}
\arg\max\limits_{\mathbf{u},\widetilde{\mathbf{u}},\mathbf{v},\mathbf{a},\mathbf{c}}
\min\limits_{l,k}\left( \mu_{k}^{l},\widetilde{\mu}_{k}^{l} \right)\\
\textrm{s.t.}\quad\sum_{l=1}^L||\mathbf{v}_k^l||^2\leq \gamma, \quad \forall k\\
\quad\quad
\mathbf{a}_{k}^l\in(\mathbb{Z}+j\mathbb{Z})^{KL};c_{k}^l\in\mathbb{Z}+j\mathbb{Z}
\end{array}\right..
\end{equation}

The above optimization problem involves complex
$\{\mathbf{u},\widetilde{\mathbf{u}},\mathbf{v}\}$ and integer
variables $\{\mathbf{a},\mathbf{c}\}$, which is a mixed integer and
continuous optimization problem.  As a result, the problem is
non-convex and requires exhaustive search\cite{liduan:2006}.

\section{Low Complexity Iterative Solution}\label{sec:alg} In this
section, we shall propose a low complexity iterative algorithm by
exploiting the special structure of the problem in
(\ref{eq:maximin}).

\subsection{Properties of the Optimal Interference Quantization Coefficients $\{\mathbf{a}^*\}$}
Although obtaining the optimal integer solution $\{\mathbf{a}^*\}$
in (\ref{eq:maximin}) is a difficult problem, we have the following
lemma to reduce the search space for $\mathbf{a}^*$.
\begin{Lem}[Properties of the Optimal
$\mathbf{a}^*$]\label{lem:integers} The optimal integer solution
$\mathbf{a}^*$ in (\ref{eq:maximin}) should belong to the following
set, i.e.,
\begin{equation}
\begin{array}{l}
(\mathbf{a}_k^l)^*\in\mathcal{A}=\{\mathbf{a}_k^l:\frac{a_i^n}{r}\not\in\mathbb{Z}+j\mathbb{Z},\forall
a_i^n\neq 0,\\
\quad\quad\quad\forall r\in(\mathbb{Z}+j\mathbb{Z})\text{ and
}|r|\neq1\}\quad \forall k,l.
\end{array}
\end{equation}
\begin{proof}
Suppose
$\{(\mathbf{v}_k^l)^*,(\mathbf{\widetilde{u}}_k^l)^*,(\mathbf{u}_k^l)^*,(c_k^l)^*,(\mathbf{a}_k^l)^*\not\in\mathcal{A}\}$
are the optimal solution. Suppose that
$\frac{(a_i^n)^*}{r}\in\mathbb{Z}+j\mathbb{Z}$, and $|r|>1$.
Although $\mathbf{a}_k^l$ will influence $\mu_{k}^{l}$ in
(\ref{eq:R_k1}) and $\widetilde{\mu}_{k}^{l}$ in (\ref{eq:R_k2}), it
is easy to verify that
$\mu_{k}^{l}(\frac{(\mathbf{u}_k^l)^*}{r},\frac{(\mathbf{a}_k^l)^*}{r})>
\mu_{k}^{l}((\mathbf{u}_k^l)^*,(\mathbf{a}_k^l)^*)$, and
$\widetilde{\mu}_{k}^{l}\bigl((c_k^l)^*r,\frac{(\mathbf{a}_k^l)^*}{r}\bigr)=
\widetilde{\mu}_{k}^{l}\bigl((c_k^l)^*,(\mathbf{a}_k^l)^*\bigr)$. In
other words, a higher data rate is achievable with the designed
parameters
$\{(\mathbf{v}_k^l)^*,(\mathbf{\widetilde{u}}_k^l)^*,\frac{(\mathbf{u}_k^l)^*}{r},(c_k^l)^*r,\frac{(\mathbf{a}_k^l)^*}{r}\}$.
As a result,
$\{(\mathbf{v}_k^l)^*,(\mathbf{\widetilde{u}}_k^l)^*,(\mathbf{u}_k^l)^*,(c_k^l)^*,(\mathbf{a}_k^l)^*\not\in\mathcal{A}\}$
cannot be an optimal solution.
\end{proof}
\end{Lem}

\subsection{Optimization of $\{\mathbf{u},\widetilde{\mathbf{u}},\mathbf{c}\}$ under fixed $\{\mathbf{v},\mathbf{a}\}$}
In this section, we fix the precoder $\mathbf{v}$ and interference
quantization coefficients $\mathbf{a}$, and we optimize the
remaining parameters
$\{\mathbf{u},\widetilde{\mathbf{u}},\mathbf{c}\}$. Note that,
$\mathbf{u}_k^l$ only influences $\mu_k^l$ in (\ref{eq:R_k1}), and
$\widetilde{\mathbf{u}}_k^l$ only influences $\widetilde{\mu}_k^{l}$
in (\ref{eq:R_k2}). For given $c_k^l$, we shall first determine the
optimal $\mathbf{u}_k^l$ and $\widetilde{\mathbf{u}}_k^l$ by
maximizing $\mu_k^{l}$ and $\widetilde{\mu}_k^{l}$, respectively,
which can be obtained by solving the following convex problems
\begin{equation}
\label{eq:decorrelator}
\begin{array}{l}
(\mathbf{u}_k^l)^*=\arg\min\limits_{\mathbf{u}_k^l}\Big(||\mathbf{u}_k^l||^2+
P\sum_{i,n}\\
\bigl||(\mathbf{u}_k^l)^H\hat{\mathbf{H}}_{ki}\mathbf{v}_i^n-a_{i}^n|+ \epsilon||\mathbf{v}_i^n||\cdot||\mathbf{u}_k^l||\bigr|^2\Big)\\
(\widetilde{\mathbf{u}}_k^l)^*=\arg\min\limits_{\widetilde{\mathbf{u}}_k^l}\Big(||\widetilde{\mathbf{u}}_k^l||^2+
P\sum_{i,n}\\
\bigl||(\widetilde{\mathbf{u}}_k^l)^H\hat{\mathbf{H}}_{ki}\mathbf{v}_i^n-c_k^la_{i}^n-1_{\{i=k\&n=l\}}|+ \epsilon||\mathbf{v}_i^n||\cdot||\widetilde{\mathbf{u}}_k^l||\bigr|^2\Big)
\end{array}.
\end{equation}

For perfect CSI ($\epsilon=0, \hat{\mathbf{H}}=\mathbf{H}$), closed
form solutions exist for (\ref{eq:decorrelator}) and the optimal
solutions are given by \cite{CF:MIMO:2009}:
\begin{eqnarray}
(\mathbf{u}_k^l)^*&=&\left(\mathbf{W}^H\mathbf{W}+\frac{1}{P}\mathbf{I}_{N\times
N}\right)^{-1}\mathbf{W}^H\boldsymbol{\alpha}_k^l, \\
(\widetilde{\mathbf{u}}_k^l)^*&=&\left(\mathbf{W}^H\mathbf{W}+\frac{1}{P}\mathbf{I}_{N\times
N}\right)^{-1}\mathbf{W}^H\boldsymbol{\beta}_k^l,
\end{eqnarray}
where
$\mathbf{W}=[\mathbf{H}_{k1}\mathbf{v}_1^1,...,\mathbf{H}_{k1}\mathbf{v}_1^L,\mathbf{H}_{k2}\mathbf{v}_2^1,...,\mathbf{H}_{kK}\mathbf{v}_K^L]^H$
is a $KL\times N$ matrix,
$\boldsymbol{\alpha}_k^l=[a_1^1,...,a_1^L,a_2^1,...,a_K^L]^H$ is a
$KL\times1$ vector, and
$\boldsymbol{\beta}_k^l=[c_k^la_1^1,...,c_k^la_1^L,c_k^la_2^1,...,1,...,c_k^la_K^L]^H$
is also a $KL\times1$ vector. On the other hand, when the CSI is
imperfect, there is no closed form solution. Since
(\ref{eq:decorrelator}) is a standard convex problem,
$(\mathbf{u}_k^l)^*$ and $(\widetilde{\mathbf{u}}_k^l)^*$ can be
obtained iteratively using an interior-point method or efficient
gradient search \cite{Convex:2004}.

Next, we shall optimize the complex integers $c_k^l$ for a given
$\widetilde{\mathbf{u}}_k^l$. The solution of $(c_k^l)^*$ for given
$(\widetilde{\mathbf{u}}_k^l)^*$ is summarized in the lemma below.
\begin{Lem} [Optimal Integer Solution of $(c_k^l)^*$ given
$(\widetilde{\mathbf{u}}_k^l)^*$] \label{lem:b,c} Given
$(\widetilde{\mathbf{u}}_k^l)^*$ in (\ref{eq:decorrelator}), the
optimal $(c_k^l)^*$ is given by:
\begin{eqnarray}
\label{eq:b_k^L}
(c_k^l)^*=\argmin_{\Re\{c_k^l\}\in\bigl[\Re\{\tau\},\Re\{\kappa\}\bigr],
\Im\{c_k^l\}\in\bigl[\Im\{\tau\},\Im\{\kappa\}\bigr]}f(c_k^l),
\end{eqnarray}
where
\begin{equation}
\tau=(\widetilde{c}_k^l)^*-(1+j) \text{ and
}\kappa=(\widetilde{c}_k^l)^*+(1+j),
\end{equation}
\begin{equation}
\begin{array}{l}
f(c_k^l)=\\
\sum\limits_{i,n}\Bigl||(\widetilde{\mathbf{u}}_k^l)^H\hat{\mathbf{H}}_{ki}\mathbf{v}_i^n-c_k^la_{i}^n-1_{\{i=k\&n=l\}}|+\epsilon||\mathbf{v}_i^n||\cdot||\widetilde{\mathbf{u}}_k^l||\Bigr|^2,
\end{array}
\end{equation}
and
\begin{equation}\label{eq:relax_c}
(\widetilde{c}_k^l)^*=\argmin_{\widetilde{c}_k^l\in\mathbb{C}}f(\widetilde{c}_k^l).
\end{equation}

\end{Lem}
\begin{proof}
$c_k^l$ only influences $\widetilde{\mu}_k^{l}$ in (\ref{eq:R_k2}).
If we relax $c_k^l\in\mathbb{Z}+j\mathbb{Z}$ to
$\widetilde{c}_k^l\in\mathbb{C}$, the function
$f(\widetilde{c}_k^l)$ is a convex function. Therefore, if
$(\widetilde{c}_k^l)^*=\arg\min_{\widetilde{c}_k^l\in\mathbb{C}}f(\widetilde{c}_k^l)$,
the optimal $c_k^l$ is one of the complex integers close to
$(\widetilde{c}_k^l)^*$, i.e.,
\begin{equation}
(c_k^l)^*=\argmin_{\Re\{c_k^l\}\in\bigl[\Re\{\tau\},\Re\{\kappa\}\bigr],
\Im\{c_k^l\}\in\bigl[\Im\{\tau\},\Im\{\kappa\}\bigr]}f(c_k^l),
\end{equation}
where $\tau=(\widetilde{c}_k^l)^*-(1+j)$,
$\kappa=(\widetilde{c}_k^l)^*+(1+j)$.
\end{proof}

As a result, given $\{\mathbf{v},\mathbf{a}\}$, we shall use the
following algorithm to optimize
$\{\mathbf{u},\widetilde{\mathbf{u}},\mathbf{c}\}$. \vspace{10pt}

\noindent
\begin{tabular}{p{8.3cm}}
\hline {\bf Subalgorithm A:} Optimization Algorithm for
$\{\mathbf{u},\widetilde{\mathbf{u}}\}$
and $\mathbf{c}$ under fixed $\{\mathbf{v},\mathbf{a}\}$ \\
\hline
 \begin{minipage}[t]{8.3cm}
\begin{itemize}
\item{\bf Step 1:} For a given realization of $\mathbf{v}$, $\mathbf{a}_k^l\in\mathcal{A}$,
initialize $\mathbf{c}(0)$. Set the iteration steps $m=0$.
\item{\bf Step 2:} For the given
$\{\mathbf{v},\mathbf{a}\}$, solve the convex optimization problem
(\ref{eq:decorrelator}) to obtain the corresponding optimal first
stage decorrelator $\mathbf{u}$.
\item{\bf Step 3:} For the given
$\{\mathbf{v},\mathbf{a},\mathbf{c}(m)\}$, solve the convex
optimization problem (\ref{eq:decorrelator}) to obtain the
corresponding optimal second stage decorrelator
$\widetilde{\mathbf{u}}(m+1)$.
\item{\bf Step 4:} For the given
$\{\mathbf{v},\mathbf{a}, \widetilde{\mathbf{u}}(m+1)\}$, obtain the
optimal positive integers $\mathbf{c}(m+1)$ by solving the
problem (\ref{eq:b_k^L}). Specifically, solve the convex problem
(\ref{eq:relax_c}) to obtain the relaxed value
$(\widetilde{c}_k^l)^*$, and then check the nearby integers around
$(\widetilde{c}_k^l)^*$ to obtain $\mathbf{c}(m+1)$ as shown in
(\ref{eq:b_k^L}).
\item{\bf Step 5:} Continue until
$\widetilde{\mathbf{u}}(m)=\widetilde{\mathbf{u}}(m+1)$, and
$\mathbf{c}(m)=\mathbf{c}(m+1)$. \vspace{5pt}
\end{itemize}
\end{minipage} \\
\hline
\end{tabular}

\vspace{10pt}

The convergence proof is given in Appendix \ref{app:alg}.


\subsection{Optimization of $\{\mathbf{v},\mathbf{a}\}$ under fixed $\{\mathbf{u},\widetilde{\mathbf{u}},\mathbf{c}\}$}
Given $\{\mathbf{u},\widetilde{\mathbf{u}},\mathbf{c}\}$, optimizing
$\{\mathbf{v},\mathbf{a}\}$ is not trivial. To obtain low complexity
solutions, we shall first relax the interference quantization
coefficients from $\mathbf{a}_k^l\in(\mathbb{Z}+j\mathbb{Z})^{KL}$
to $\mathbf{a}_k^l\in\mathbb{C}^{KL}$. Define
\begin{equation}
\begin{array}{l}
g^{l}_k(\mathbf{v},\mathbf{a}_k^l)=||\mathbf{u}_k^l||^2+
P\sum_{i,n}\\
\Bigl||(\mathbf{u}_k^l)^H\hat{\mathbf{H}}_{ki}\mathbf{v}_i^n-a_{i}^n|+\epsilon||\mathbf{v}_i^n||\cdot||\mathbf{u}_k^l||\Bigr|^2\\
\widetilde{g}^{l}_k(\mathbf{v},\mathbf{a}_k^l)=||\widetilde{\mathbf{u}}_k^l||^2+
P\sum_{i,n}\\
\Bigl||(\widetilde{\mathbf{u}}_k^l)^H\hat{\mathbf{H}}_{ki}\mathbf{v}_i^n-c_k^la_{i}^n-1_{\{i=k\&n=l\}}|+\epsilon||\mathbf{v}_i^n||\cdot||\widetilde{\mathbf{u}}_k^l||\Bigr|^2\nonumber,
\end{array}
\end{equation}
where the right hand side of these two equations are the denominator
of (\ref{eq:R_k1}) and (\ref{eq:R_k2}), respectively.
$\{g^{l}_k,\widetilde{g}^{l}_k\}$ are convex functions w.r.t.
$\{\mathbf{v},\mathbf{a}_k^l\}$ for all $\{l,k\}$. Given
$\{\mathbf{u},\widetilde{\mathbf{u}},\mathbf{c}\}$, the max-min
problem in (\ref{eq:maximin}) is equivalent to
\begin{eqnarray}
\label{eq:w_and_a}
&\max_{\mathbf{v},\mathbf{a},t}& t \nonumber\\
&\textrm{s.t.}& t\leq \mu_{k}^{l}\text{ and } t\leq \widetilde{\mu}_{k}^{l},\quad\forall k,l\nonumber\\
&&\sum_{l=1}^L||\mathbf{v}_k^l||^2\leq \gamma,\forall k,
\end{eqnarray}
which is not a convex problem, because
$\log\left(g^{l}_k(\mathbf{v},\mathbf{a}_k^l)\right)$ and
$\log\left(\widetilde{g}^{l}_k(\mathbf{v},\mathbf{a}_k^l)\right)$
are not convex functions w.r.t. $\{\mathbf{v},\mathbf{a}_k^l\}$ for
all $\{l,k\}$. However, we show that (\ref{eq:w_and_a}) is
equivalent to a convex problem given in the following lemma.
\begin{Lem}[Equivalent Convex Problem]
\label{lem:w_and_a_convex} The problem in (\ref{eq:w_and_a}) is
equivalent to the following convex problem
\begin{eqnarray}
\label{eq:w_and_a_convex}
&\min_{\mathbf{v},\mathbf{a},t}& t \nonumber\\
&\textrm{s.t.}& g^{l}_k(\mathbf{v},\mathbf{a}_k^l)\leq t \text{ and } \widetilde{g}^{l}_k(\mathbf{v},\mathbf{a}_k^l)\leq t, \forall l,k\nonumber\\
&&\sum_{l=1}^L||\mathbf{v}_k^l||^2\leq \gamma,\forall k.
\end{eqnarray}
\begin{proof}
We show that given
$\{\mathbf{u},\widetilde{\mathbf{u}},\mathbf{c}\}$, the max-min
problem in (\ref{eq:maximin}) is equivalent to
\begin{eqnarray}
\arg\min_{\mathbf{v},\mathbf{a}}\max_{l,k}(g^{l}_k,\widetilde{g}^l_k).
\end{eqnarray}

First, note that given $\{\mathbf{v},\mathbf{a}\}$, the optimal
$\{l^*,k^*\}$ obtained from
$\arg\max_{l,k}(g^{l}_k,\widetilde{g}^l_k)$ is the same as
$\arg\min_{l,k}(\mu_k^l,\widetilde{\mu}_k^l)$. Furthermore, note
that
$\arg\min_{\mathbf{v},\mathbf{a}}(g^{l^*}_{k^*},\widetilde{g}^{l^*}_{k^*})=\arg\max_{\mathbf{v},\mathbf{a}}(\mu^{l^*}_{k^*},\widetilde{\mu}^{l^*}_{k^*})$.

Therefore,
\begin{equation}
\label{eq:maximin=minimax}
\arg\min_{\mathbf{v},\mathbf{a}}\max_{l,k}(g^{l}_k,\widetilde{g}^l_k)=\arg\max_{\mathbf{v},\mathbf{a}}\min_{l,k}(\mu^{l}_k,\widetilde{\mu}^l_k).
\end{equation}

Similarly, we can also verify that (\ref{eq:maximin=minimax}) is
equivalent to (\ref{eq:w_and_a_convex}).
\end{proof}
\end{Lem}

As a result, given
$\{\mathbf{u},\widetilde{\mathbf{u}},\mathbf{c}\}$, we shall use the
following algorithm to optimize $\{\mathbf{v},\mathbf{a}\}$ by using
an interior-point method\cite[Chap.11]{Convex:2004}.
\vspace{10pt}

\noindent
\begin{equation}\nonumber
\begin{array}{l}
\hline \text{{\bf Subalgorithm B:} Optimization Algorithm for }
 \{\mathbf{v},\mathbf{a}\} \\ \text{ under fixed } \{\mathbf{u},\widetilde{\mathbf{u}},\mathbf{c}\}
 \quad\quad\quad\quad\quad\quad\quad\quad\quad\quad\quad\quad\quad\quad\quad\quad
 \\\hline
\end{array}
\end{equation}
\begin{itemize}
\item{\bf Step 1:} For a given realization of $\{\mathbf{u},\widetilde{\mathbf{u}},\mathbf{c}\}$,
initialize $q>0$, $\{\mathbf{v}_0,\mathbf{a}_0\}$, $\nu>1$ and
tolerance $\varsigma>0$.
\item{\bf Step 2:} Solve the unconstrained convex problem
$\{\mathbf{v}^*(q),\mathbf{a}^*(q)\}=\arg\min_{\mathbf{v},\mathbf{a}}qt-
\sum_{k,l}\big(\log(t-g^{l}_k(\mathbf{v},\mathbf{a}_k^l))+\log(t-\widetilde{g}^{l}_k(\mathbf{v},\mathbf{a}_k^l))\big)$,
starting at $\{\mathbf{v}_0,\mathbf{a}_0\}$ by efficient gradient
search\footnote{ For example, using the Newton's method
\cite[Alg 9.5]{Convex:2004}.}.
\item{\bf Step 3:} If $\frac{1}{q}<\varsigma$, stop and set
$\{\mathbf{v}^*,\mathbf{a}^*\}=\{\mathbf{v}^*(q),\mathbf{a}^*(q)\}$,
otherwise go to the next step.
\item{\bf Step 4:} Set
$\{\mathbf{v}_0,\mathbf{a}_0\}=\{\mathbf{v}^*(q),\mathbf{a}^*(q)\}$,
\item{\bf Step 5:} Set $q=\nu q$,
and go to the step 2. \vspace{5pt}
\end{itemize}
\vspace{-10pt}
\begin{equation}\nonumber
\begin{array}{l}
\hline
\quad\quad\quad\quad\quad\quad\quad\quad\quad\quad\quad\quad\quad\quad\quad\quad\quad\quad\quad
\quad\quad\quad\quad\quad
\end{array}
\end{equation}

The convergence proof is shown in Appendix \ref{app:alg}.

\subsection{A Summary of the Overall Solution}
The top-level optimization algorithm is summarized below (and
illustrated in Fig. \ref{fig:algorithm_diagram}): \vspace{5pt}

\begin{figure}
\centering
\includegraphics[width = 6cm]{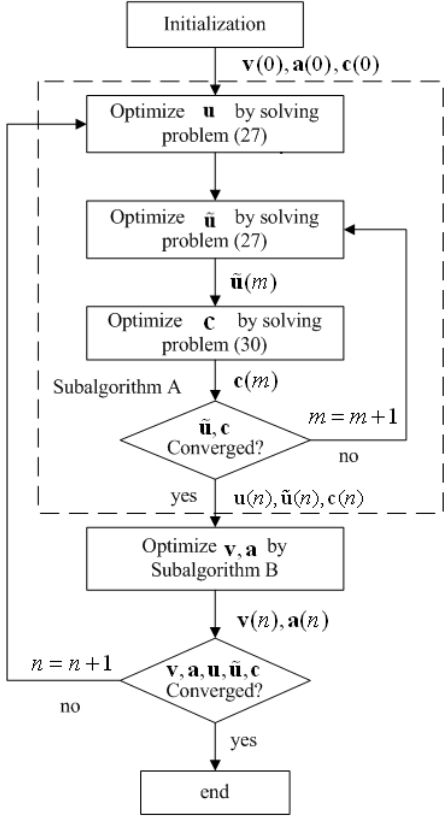}
\caption{ Illustration of the top-level optimization algorithm.
Specifically, $\mathbf{v}$ denotes the precoders, $\mathbf{a}$
denotes the interference quantization coefficients, $\mathbf{u}$
denotes the first stage decorrelators, $\widetilde{\mathbf{u}}$
denotes the second stage decorrelators, and $\mathbf{c}$ denotes the
scaling coefficients.} \label{fig:algorithm_diagram}
\end{figure}

\noindent
\begin{tabular}{p{8.3cm}}
\hline {\bf Algorithm 1:} Top-Level Optimization
Algorithm\label{alg:v,a} \\
\hline
 \begin{minipage}[t]{8.3cm}
\begin{itemize}
\item{\bf Step 1:} Initialize $\mathbf{v}_k^l(0)$, $\mathbf{a}_k^l(0)\in\mathcal{A}$
and $c_k^l(0)$. Set the iteration steps $m=0$.
\item{\bf Step 2:} For the given $\{\mathbf{v}(m),\mathbf{a}(m)\}$,
apply the subalgorithm A to obtain $\{\mathbf{u}(m+1),
\widetilde{\mathbf{u}}(m+1),\mathbf{c}(m+1)\}$.
\item{\bf Step 3:} For the obtained $\{\mathbf{u}(m+1),
\widetilde{\mathbf{u}}(m+1),\mathbf{c}(m+1)\}$, apply subalgorithm B
to obtain the corresponding optimal
$\{\mathbf{v}(m+1),\mathbf{a}(m+1)\}$.
\item{\bf Step 4:} Continue until
$\widetilde{\mathbf{u}}(m)=\widetilde{\mathbf{u}}(m+1)$,
$\mathbf{u}(m)=\mathbf{u}(m+1)$, $\mathbf{c}(m)=\mathbf{c}(m+1)$,
$\mathbf{v}(m)=\mathbf{v}(m+1)$, and
$\mathbf{a}(m)=\mathbf{a}(m+1)$. \vspace{5pt}
\end{itemize}
\end{minipage} \\
\hline
\end{tabular}

\vspace{10pt}

The convergence proof of Algorithm 1 is shown in Appendix
\ref{app:alg}. Finally, suppose $(\mathbf{\hat{a}}_k^l)^*$ is the
solution to Algorithm 1, we could quantize
$(\mathbf{\hat{a}}_k^l)^*$ to the nearest complex integers
$(\mathbf{a}_k^l)^*=\{\{(a_i^n)^*\}_{i=1}^K\}_{n=1}^L$ such that
\begin{equation}
\label{eq:a_i^n} (a_i^n)^* =
\argmin_{a_i^n\in\mathbb{Z}+j\mathbb{Z}}|a_i^n-(\hat{a}_i^n)^*|,\forall
i,n.
\end{equation}

\begin{Rem} {\em Qualitative Comparison of the Algorithm with Conventional Approaches):}
\end{Rem}
\begin{itemize}
\item{\bf (a) Complexity comparisons:} The proposed method has a
complexity that is close to the conventional Distributive IA method
using alternating optimization \cite{IA:distributed:2008}
but as expected converges slower as seen in Table
\ref{tab:sim_time}. This is because the subproblems to be solved in
each step of the proposed algorithm (e.g. Problem
(\ref{eq:decorrelator}) and Problem (\ref{eq:w_and_a_convex})) are
convex and hence, there exists efficient algorithms (such as
interior-point method \cite{Convex:2004}).

\item{\bf (b) Limitations of other existing approaches:} For the conventional IA method \cite{IA:conventional:2008}, there is
always a feasibility problem for quasi-static interference channels
for $K>3$. All the other baselines (see Section \ref{sec:sim}) have
poor performance when $K>2$ even when the CSI is perfect.

\item{\bf (c) Imperfect CSI considerations:} Furthermore, none of the baselines
consider imperfect CSI while the proposed method takes into account
imperfect CSI in the design. Hence, they have very different
performance in the presence of imperfect CSI. ~\hfill\IEEEQED
\end{itemize}

\begin{table*}
\begin{center}
\caption{ Comparison of matlab simulation time and the
converged performance for a channel realization at transmit
SNR=$1.5$\MakeLowercase{d}B for $M=N=2$ and $L=1$ with perfect CSI
$\epsilon = 0$}
\begin{tabular}{|c|c|c|c|c|}
\hline Number of Users & Methods & Time (s) & Worst-Case Goodput (b/s/Hz) & Sum Goodput (b/s/Hz) \\
\hline \multirow{2}*{$K=3$} & Proposed Method          & 4.442407  &
1.4864 & 4.4593 \\ \cline{2-5}
& Distributive IA          & 1.579973  & 0.3306     & 2.4724 \\
\cline{2-5} \hline

\multirow{2}*{$K=4$} & Proposed Method          & 9.009197  & 0.9036
& 3.6144 \\ \cline{2-5}
& Distributive IA          & 2.995086  & 0.2537     & 2.7012 \\
\hline
\end{tabular}
\label{tab:sim_time}
\end{center}
\end{table*}

\section{Quantitative Analysis of the Proposed Robust Lattice
Alignment Solution under Symmetric Interference
Channels}\label{sec:interpretation} In this section, we shall
illustrate analytically the potential benefits of the proposed
robust lattice alignment solution versus the brute-force two-stage
ML decoding with Gaussian random inputs. To obtain a first order
comparison, we consider a {\em symmetric interference channel } as
in \cite{Symmetric:2008}, where each user has one antenna and each
transmitter tries to transmit one data stream for the desired
receiver. The symmetric channel model is as follows:
\begin{equation}
\label{eq:symmetric_CSI} \mathbf{y}_k=\mathbf{x}_k+h\sum_{i=1,i\neq
k}^K\mathbf{x}_i+\mathbf{z}_k,
\end{equation}
where we assume that the interference channel coefficients
$h\in\mathbb{Z}+j\mathbb{Z}$. In \cite{Symmetric:2008}, only real
channel coefficients are considered.

\subsection{Performance of the Proposed Method for Symmetric
Interference Channels} Since all channels are symmetric and $M=1$,
we have $(v_k)^* = 1$ for all $k$, and we can consider user 1
without loss of generality. During stage I decoding, the optimal
interference quantization coefficients are given by
$\mathbf{a}^*=[0,1,1,...,1]^H$ from Lemma \ref{lem:integers}. As a
result, we would decode the {\em aggregate interference}
$\left[\mathbf{x}_2+\mathbf{x}_3+...+\mathbf{x}_K\right]\text{ Mod
}\Lambda$ without any quantization approximation. By solving the
optimization problem in (\ref{eq:decorrelator}), we can obtain the
optimal first stage decorrelator,
\begin{equation}
(\mathbf{u})^*=\Big(\mathbf{W}^H\mathbf{W}+\frac{1}{P}\Big)^{-1}\mathbf{W}^H\mathbf{a}^*
=\frac{(K-1)h}{1+(K-1)|h|^2+1/P},
\end{equation}
where $\mathbf{W}=[1,h,\cdots,h]^H$, and hence we have
\begin{equation}
R_2^*=R_3^*...=R_K^*=\log\left(\frac{1+[(K-1)|h|^2+1]P}{K-1+(K-1)P}\right).
\end{equation}

During the second stage decoding, we have $c_k^*=h$ and
$(\widetilde{\mathbf{u}}_k^l)^*=1$, where the data rate for user 1
is given by $R_1^*=\log(P)$. As a result, the minimal achievable
data rate is given by:
\begin{equation}
\label{eq:ours}
R_{\min}=\min\left(\log(P),\log\left(\frac{1+[(K-1)|h|^2+1]P}{K-1+(K-1)P}\right)\right).
\end{equation}

As shown from (\ref{eq:ours}), the performance bottleneck of the
proposed method is the stage I decoding at the medium SNR regime.
Furthermore, the proposed method achieves similar performance as in
\cite{Symmetric:2008} and hence, the proposed method is backward
compatible with that in \cite{Symmetric:2008} under the same
symmetric interference channel realizations.

%
%

\subsection{Comparison with Brute-Force Two Stage ML Decoding with Gaussian
Inputs}\label{subsec:ML} In this section, we shall compare the
performance of the proposed method with a baseline, namely the
brute-force two-stage ML decoding with random Gaussian inputs.
Specifically, each transmitter transmits random Gaussian signals
$\mathbf{x}_k$ and each receiver of the $K$-user symmetric
interference channel in (\ref{eq:symmetric_CSI}) performs
brute-force two-stage ML decoding. Without loss of generality, we
consider user 1 in the analysis. During the first stage decoding,
the Gaussian interferences $\{\mathbf{x}_2,...,\mathbf{x}_K \}$ are
decoded by using brute-force ML. The decoded aggregate interference
is canceled from the received signal $\mathbf{y}_1$ and the desired
signal $\mathbf{x}_1$ is decoded by a second stage ML decoding. Note
that unlike the proposed method, random Gaussian signals are
transmitted and hence, the interference space is random. In order to
have successful decoding in both stages of this system, the data
rate has to satisfy the following constraints \cite{Cover:06}:
\begin{eqnarray}
R_1&\leq&\log\left(1+P\right)\nonumber\\
\sum\nolimits_{k=2}^KR_k&\leq&\log\left(1+\frac{(K-1)P|h|^2}{1+P}\right).
\end{eqnarray}
Therefore, the minimal achievable data rate for all the users is
given by:
\begin{equation}
\begin{array}{l}
R_{B2}=\\
\min\Big(\frac{1}{(K-1)}\log\big(1+\frac{(K-1)P|h|^2}{1+P}\big),
\log(1+P) \Big),
\end{array}
\end{equation}
where the bottleneck is also to decode the interference at medium
SNR regime. Similarly, the conditions in which the proposed method
offers a performance gain over the baseline system is given by:
\begin{equation}
\begin{array}{ll}
&\log\left(\frac{1+[(K-1)|h|^2+1]P}{K-1+(K-1)P}\right)\\
>&\frac{1}{(K-1)}\log\left(1+\frac{(K-1)|h|^2P}{1+P}\right).
\end{array}
\end{equation}
The condition reduces to $|h|^2\geq (1+\frac{1}{P})$ for large $K$
and $|h|^2\geq \frac{(K-1)^{\frac{K-1}{K-2}}-1}{K-1}$ for medium
SNR. In other words, only mild conditions are needed for a
performance gain over the baseline system.

\section{Performance of Proposed Method under $K$-user
Interference Channels}\label{sec:sim}
\subsection{Comparison with Conventional Interference Alignment
Solution}As pointed out in Section \ref{sec:sub_IA}, conventional IA
can achieve $\frac{3M}{2}$ DoF in the 3-user quasi-static MIMO
interference channels with $M>1$ antennas for all transmitters and
receivers\cite{IA:conventional:2008}. However, it requires
full rank channel matrices whose coefficients are randomly drawn
from a continuous distribution\footnote{Therefore, the special
cases are not considered in conventional IA. e.g., key hole effect,
line-of-sight and closely spaced antennas.}, as well as the
feasibility condition. On the other hand, our proposed method works
for any channel realizations and does not have feasibility problem.
In this section, we shall illustrate that the proposed method could
achieve the same DoF as the conventional
IA\cite{IA:conventional:2008} if it is feasible. Without loss of
generality, consider the case when $M$ is even and denote the
solution of the alignment method in \cite{IA:conventional:2008} by
\begin{equation}
\begin{array}{lll}
\mathbf{V}_k(\text{IA})&=&[\mathbf{v}_k^1(\text{IA}),...,\mathbf{v}_k^{M/2}(\text{IA})],\\
\mathbf{U}_k(\text{IA})&=&[\mathbf{u}_k^1(\text{IA}),...,\mathbf{u}_k^{M/2}(\text{IA})],
\end{array}
\end{equation}
where $\mathbf{v}_k^l(\text{IA})$ and $\mathbf{u}_k^l(\text{IA})$
are the precoder and decorrelator for the $l$-th data stream of user
$k$ respectively, such that
\begin{equation}
\begin{array}{l}
\mathbf{u}_k^l(\text{IA})\mathbf{H}_{kk}\mathbf{v}_k^l(\text{IA})\neq0;\\
\mathbf{u}_k^l(\text{IA})\mathbf{H}_{ki}\mathbf{v}_i^n(\text{IA})=0,\forall
i\neq k \text{ or } \forall n\neq l.
\end{array}
\end{equation}

 The same DoF of $\frac{3M}{2}$ can be
achieved in our proposed method by setting $L=\frac{M}{2}$ and
choosing the design parameters as
\begin{equation}
\label{eq:IA_uv}
\begin{array}{l}
\mathbf{a}_k^l=\mathbf{0};\mathbf{u}_k^l=\mathbf{0};c_k^l=1;\mathbf{v}_k^l=\mathbf{v}_k^l(\text{IA});\\
\widetilde{\mathbf{u}}_k^l=\frac{P\zeta}{||\mathbf{u}_k^l(\text{IA})||^2+P|\zeta|^2}\mathbf{u}_k^l(\text{IA}),
\end{array}
\end{equation}
where
$\zeta=\big(\mathbf{u}_k^l(\text{IA})\big)^H\mathbf{H}_{kk}\mathbf{v}_k^l(\text{IA})\neq
0$. Since $\mathbf{a}_k^l=\mathbf{0}$, there is no stage I decoding
and hence, from (\ref{eq:R_k1}), the data rate constraint on the
data stream is $\infty$. This means that there is no data rate
constraint associated with stage I decoding. As a result, the
desired data stream $\mathbf{x}_k^l$ is decoded directly at stage
II. From (\ref{eq:R_k2}) and (\ref{eq:IA_uv}), the achievable data
rate $R_k^l$ for data stream $\mathbf{x}_k^l$ is given by:
\begin{equation}
R_k^l<\log\left(1+\frac{P|\zeta|^2}{||\mathbf{u}_k^l(\text{IA})||^2}\right).
\end{equation}

Therefore, the DoF $3L=\frac{3M}{2}$ can also be achieved in our
proposed method. Hence, our proposed method achieves the same DoF
performance as the conventional interference alignment method in
\cite{IA:conventional:2008,IA:M*N_MIMO:2008}, as long as the problem
is feasible. On the other hand, when the conventional alignment
method is infeasible, our proposed solution still offers significant
performance gains compared with various baseline systems.

\subsection{Numerical Simulations}
In this section, we shall compare the system goodput (b/s/Hz
successfully received) of the proposed robust lattice alignment
method (designed to optimize the worst-case data rate) with five
baselines. Baseline 1 refers to the TDMA method. Baseline 2 refers
to the brute-force two-stage ML decoding with Gaussian inputs
described in Section \ref{subsec:ML}. Baseline 3 refers to the
generalized HK method for $K>2$ users, where each user's message is
split into one private part and several common parts as described in
\cite{Threeusers:2008}. Baseline 4 refers to distributive IA based
on the alternating optimization method in
\cite{IA:distributed:2008}, which tries to minimize the sum leakage
interference. Baseline 5 refers to the conventional IA method in
\cite{IA:conventional:2008}. In the simulation, we assume that all
the channel coefficients are i.i.d. zero mean unit variance
circularly symmetric complex Gaussian.

\begin{figure}
\begin{center}
  \subfigure[Perfect CSI (CSI error $\epsilon=0$) ]
  {\resizebox{7.8cm}{!}{\includegraphics{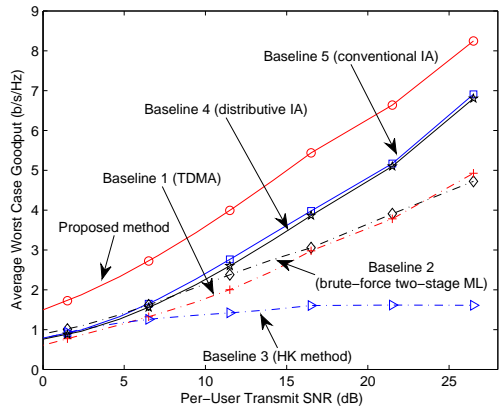}}}
  \subfigure[Imperfect CSI (CSI error $\epsilon=0.1$) ]
  {\resizebox{7.8cm}{!}{\includegraphics{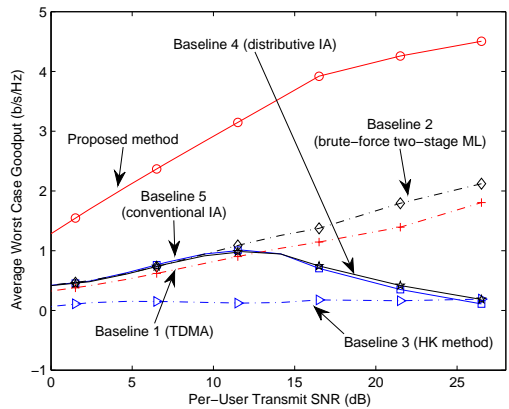}}}
  \end{center}
    \caption{Comparison of average worst-case goodput {(b/s/Hz
successfully received)} versus transmit SNR (dB) with CSI errors
$\epsilon = \{0, 0.1\}$. The setup is given by $K = 3$ (number of
users), $M=N = 2$ (transmit and receive antennas) and $L=1$ (number
of data stream).
 }
    \label{fig:3user_IA_worst_case}
\end{figure}

\begin{figure}
\begin{center}
  \subfigure[Perfect CSI (CSI error $\epsilon=0$)]
  {\resizebox{7.8cm}{!}{\includegraphics{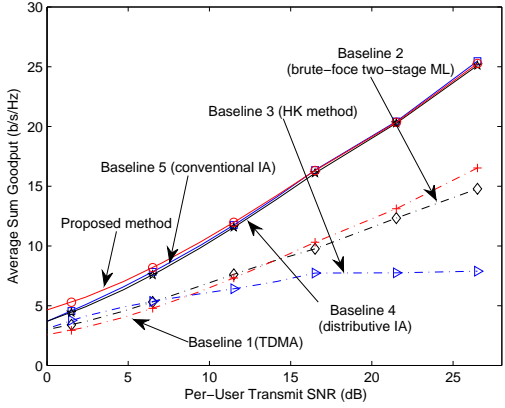}}}
  \subfigure[Imperfect CSI (CSI error $\epsilon=0.1$) ]
  {\resizebox{7.8cm}{!}{\includegraphics{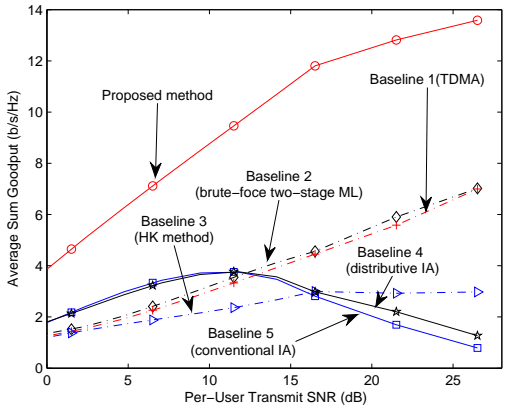}}}
  \end{center}
    \caption{Comparison of the average sum goodput (b/s/Hz
successfully received) versus transmit SNR (dB) with CSI errors
$\epsilon = \{0,0.1\}$. The setup is the same as that of Fig.
\ref{fig:3user_IA_worst_case}.
}
    \label{fig:3user_IA_sum}
\end{figure}

Fig. \ref{fig:3user_IA_worst_case} and Fig. \ref{fig:3user_IA_sum}
illustrate the average worst-case goodput and the sum goodput versus
transmit SNR (dB), respectively, at CSI errors $\epsilon=\{0,0.1\}$
for $K=3$, $M=N=2$ and $L=1$. It can be observed that our proposed
method outperforms all the baselines under both perfect and
imperfect CSI. Furthermore, the conventional IA and distributive IA
methods achieve good performance with perfect CSI but they are very
sensitive to CSI errors, especially in the high SNR region. On the
other hand, our proposed method is robust to CSI errors and
outperforms all the baselines as illustrated in Fig.
\ref{fig:3user_IA_worst_case}. Note that our optimization
objective is chosen to be the worst-case data rate for fairness
consideration, which may cause some penalty\footnote{Had we changed
the optimization objective to sum rate, the proposed method would
also outperform the others in Fig. \ref{fig:3user_IA_sum}(a).} on
the system sum rate (as illustrated in Fig.
\ref{fig:3user_IA_sum}(a)).

\begin{figure}
\begin{center}
  \subfigure[Perfect CSI (CSI error $\epsilon=0$)]
  {\resizebox{7.8cm}{!}{\includegraphics{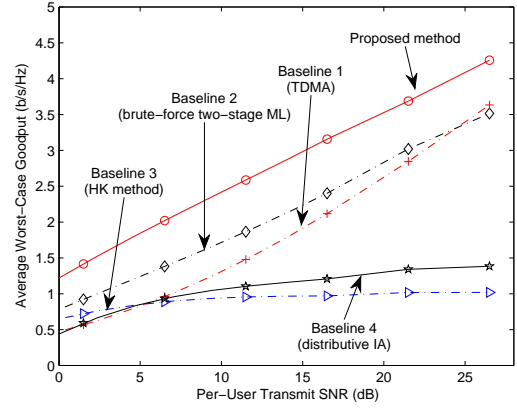}}}
  \subfigure[Imperfect CSI (CSI error $\epsilon=0.1$)]
  {\resizebox{7.8cm}{!}{\includegraphics{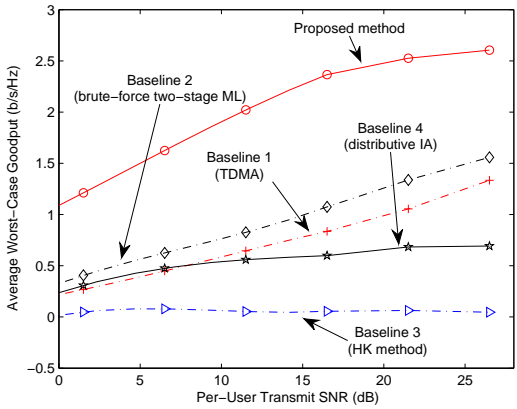}}}
  \end{center}
    \caption{Comparison of average worst-case goodput (b/s/Hz
successfully received) versus transmit SNR (dB) with CSI errors
$\epsilon = \{0, 0.1\}$. The setup is given by $K = 4$ (number of
users), $M=N = 2$ (transmit and receive antennas) and $L=1$ (number
of data stream). Note that in this configuration, the conventional
IA method is not feasible\cite{Feasibility:MIMO:2009}.
}
    \label{fig:4user_worst}
\end{figure}

Fig. \ref{fig:4user_worst} illustrates the average worst-case
goodput versus transmit SNR (dB) with CSI errors $\epsilon =
\{0,0.1\}$ for $K=4$, $M=N=2$ and $L=1$. Note that in this
configuration, the conventional IA method
\cite{IA:conventional:2008} is not feasible but the proposed method
could achieve significant performance gain over all the other
baselines.

\begin{figure}
\begin{center}
  \subfigure[Average worst-case goodput]
  {\resizebox{7.8cm}{!}{\includegraphics{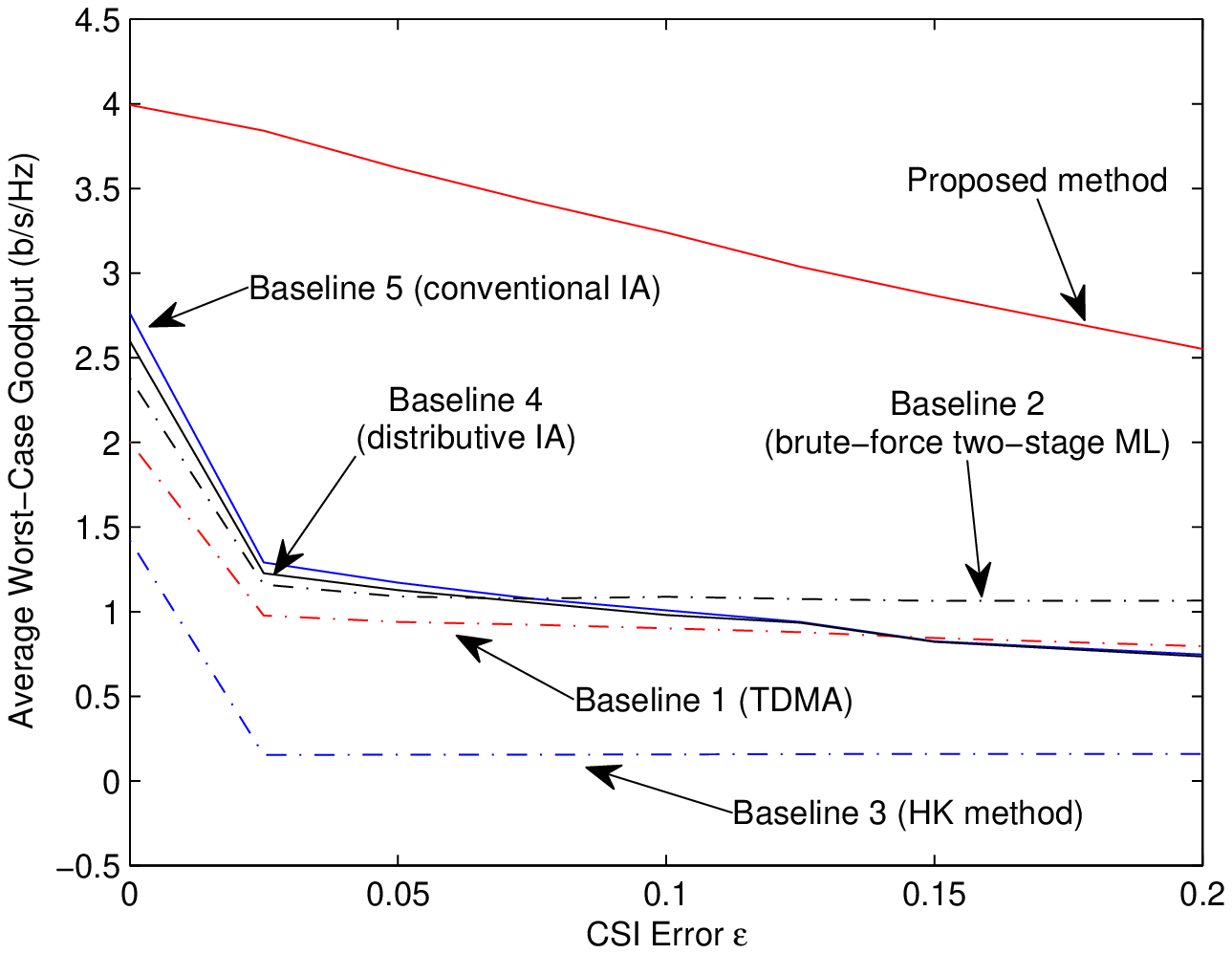}}}
  \subfigure[Average sum goodput]
  {\resizebox{7.8cm}{!}{\includegraphics{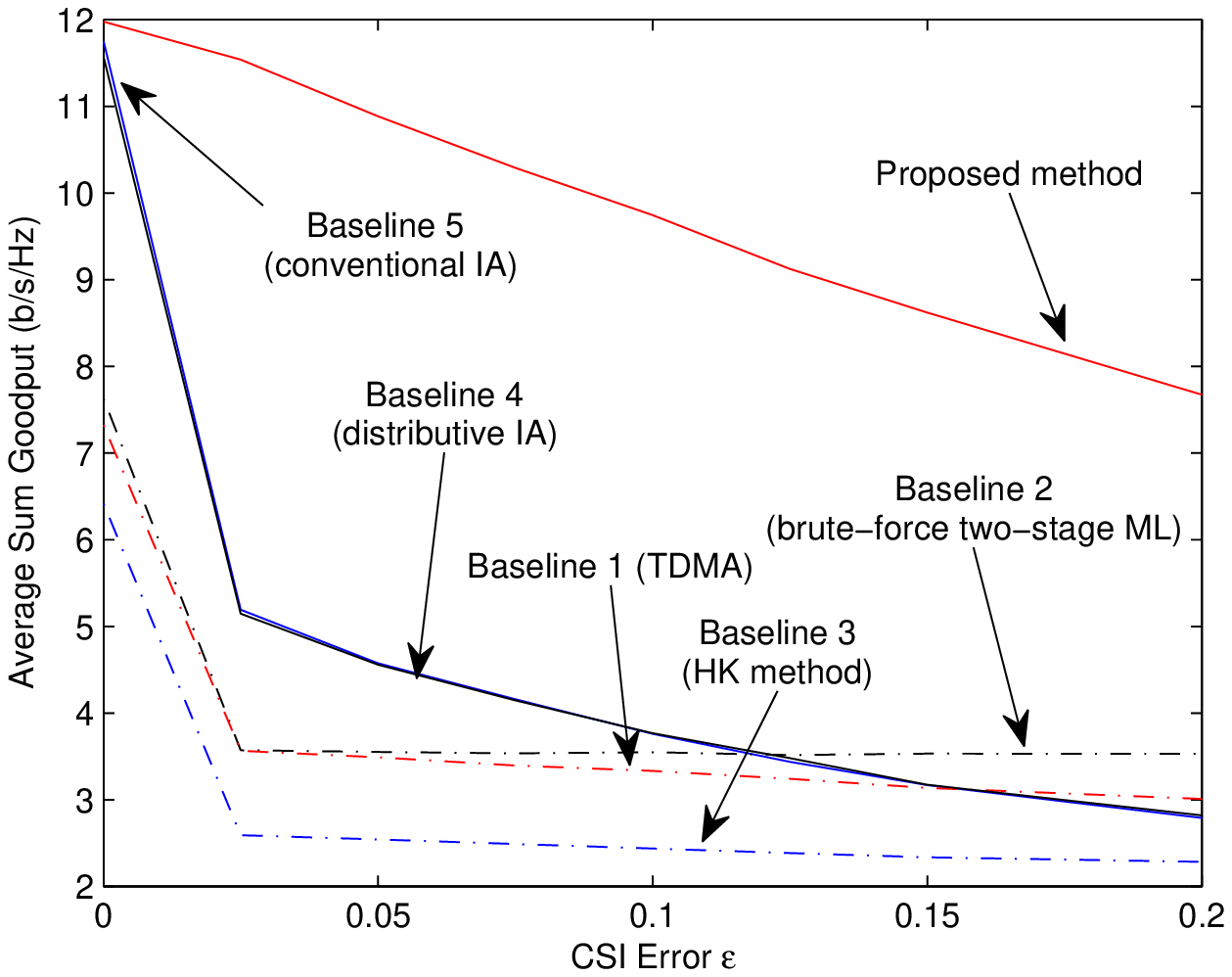}}}
  \end{center}
    \caption{Comparison of average sum and worst-case goodput (b/s/Hz
successfully received) versus CSI error $\epsilon$ at transmit
SNR=11.5dB. The setup is given by $K = 3$ (number of users), $M=N =
2$ (transmit and receive antennas) and $L=1$ (number of data
stream). }
    \label{fig:3users_csi_err}
\end{figure}

\begin{figure}
\centering
\includegraphics[width = 8.5cm]{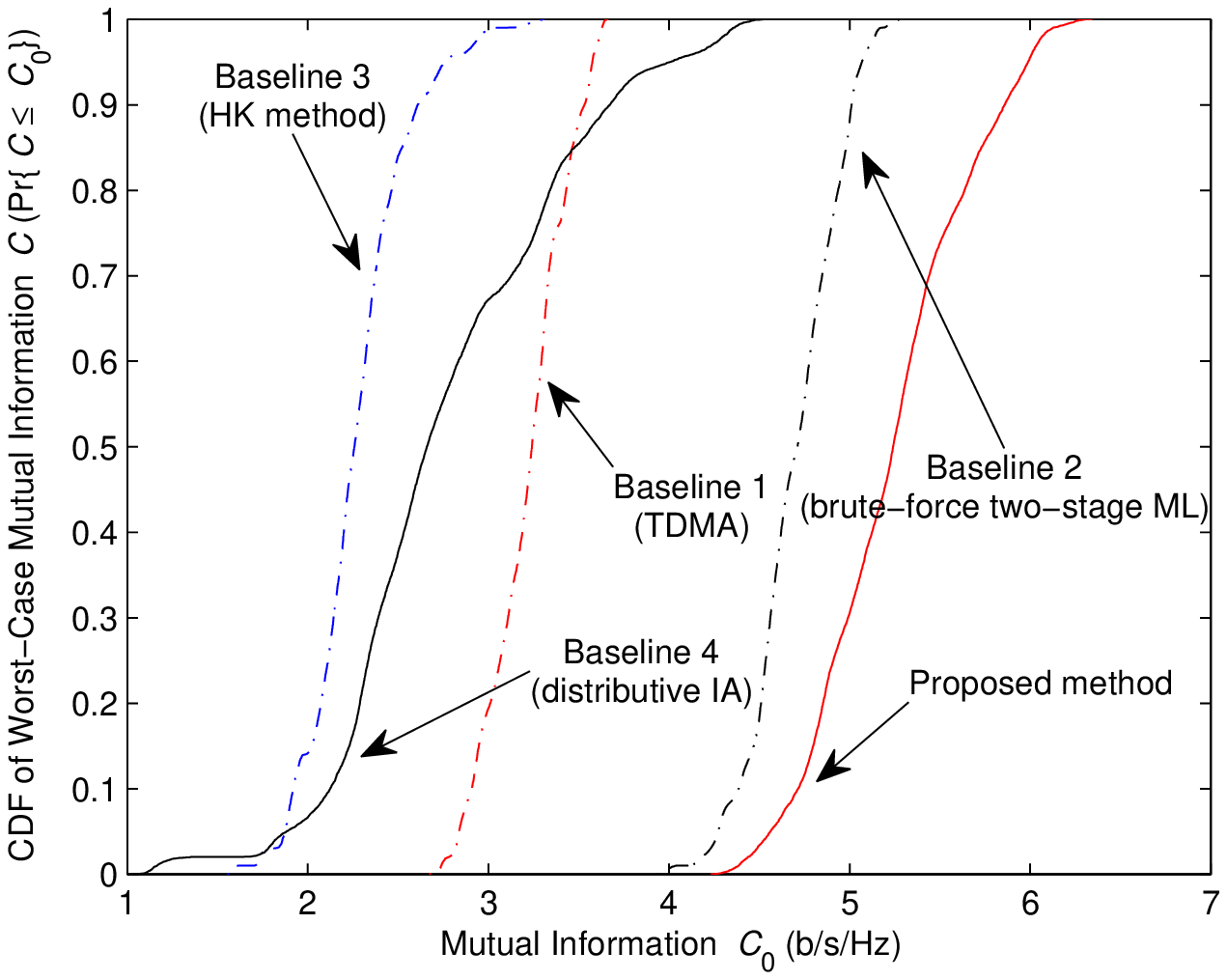}
\caption{The cumulative distribution function (CDF) of the
instantaneous worst-case mutual information per user at the transmit
SNR=11.5dB for $K=4$, $M=N=4$ and $L=2$ with CSI error $\epsilon =
0.1$.} \label{fig:cdf_worst_4users}
\end{figure}

Fig. \ref{fig:3users_csi_err} illustrates the average sum and
worst-case goodput versus CSI error $\epsilon$ at transmit
SNR=11.5dB for $K=3$, $M=N=2$ and $L=1$. It can be observed that all
the baselines are very sensitive to imperfect CSI. Even a small CSI
error will result in a significant degradation in the system
performance. On the other hand, our method is quite robust to
imperfect CSI. Fig. \ref{fig:cdf_worst_4users} illustrates the
cumulative distribution function (CDF) of the worst-case
instantaneous mutual information per user at the transmit SNR=11.5dB
for $K=4$, $M=N=4$ and $L=2$ with CSI error $\epsilon = 0.1$.
Observe that our proposed method still offers significant
performance gain in the outage sense.

\begin{figure}
\centering
\includegraphics[width = 8.5cm]{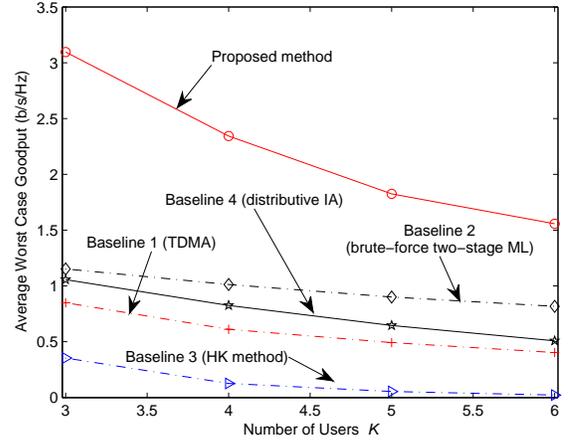}
\caption{ The average worst-case goodput per user versus the
number of users $K$ at transmit SNR=1.5dB for $M=N=4$ and $L=2$ with
CSI error $\epsilon = 0.1$.} \label{fig:number_users_e_0p1}
\end{figure}

\begin{figure}
\centering
\includegraphics[width = 8.5cm]{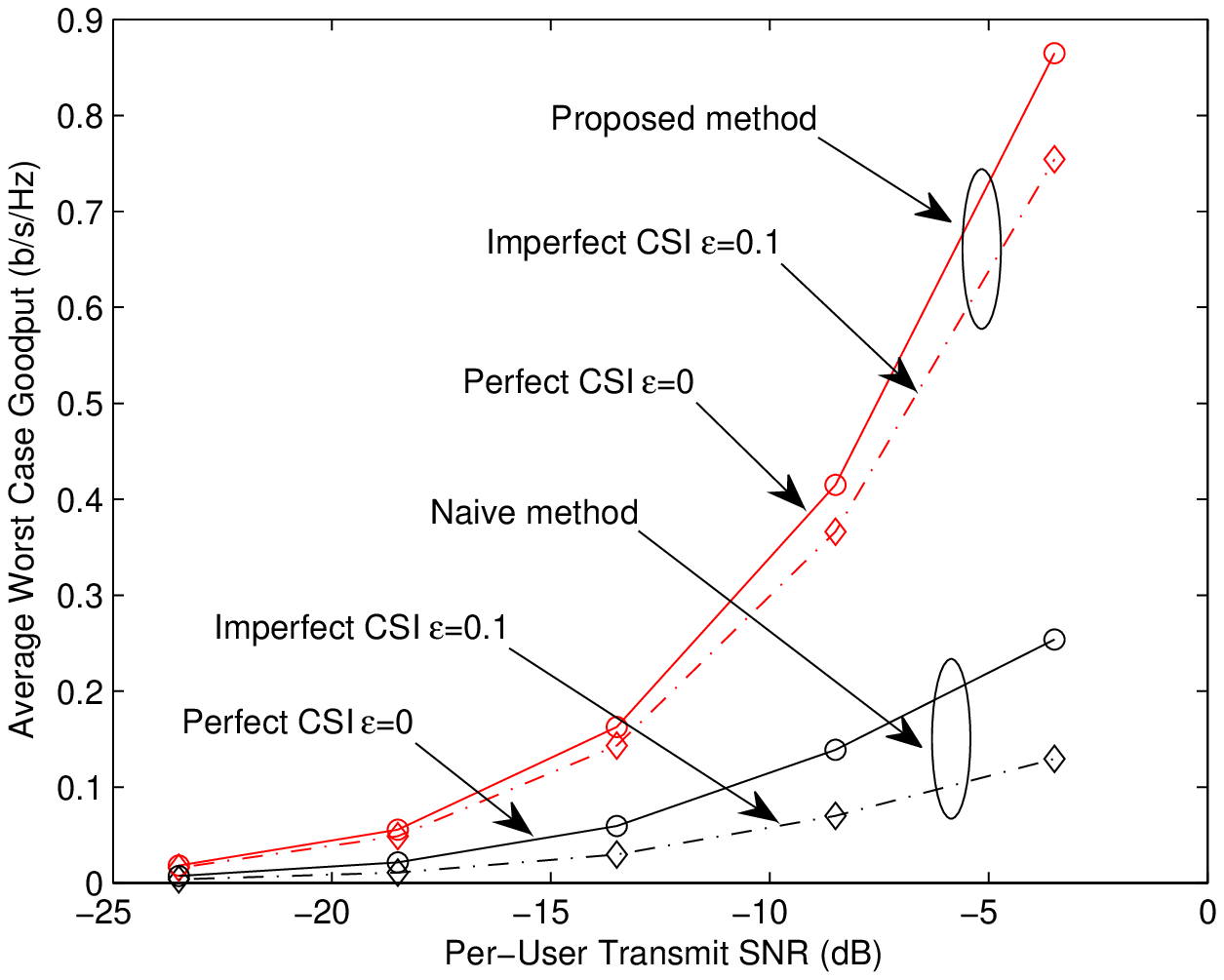}
\caption{Illustration of worst-case goodput (b/s/Hz successfully
received) at low SNR with CSI errors $\epsilon = \{0,0.1\}$ for
$K=4$, $M=N=2$ and $L=1$. The naive method refers to treating
interference as noise without requiring any CSI at the transmitter.
At low SNR, treating interference as noise is a reasonable strategy
\cite{IC:gaussian:capacity} and is also a robust method as no CSIT
is needed. } \label{fig:low_SNR_worst}
\end{figure}

Fig. \ref{fig:number_users_e_0p1} illustrates the average worst-case
goodput per user versus the number of users $K$ at transmit
SNR=$1.5$dB for $M=N=4$ and $L=2$ with CSI error $\epsilon = 0.1$.
Observe that the performance degrades as $K$ increases, and it is
more difficult to decode the desired signal. Nevertheless, the
proposed method outperforms all the baselines at all $K$. Finally,
we compare the performance of the proposed method at low SNR with a
naive method that treats interference as noise\footnote{It is shown
in \cite{IC:gaussian:capacity} that treating interference as noise
is a reasonable strategy at low SNR. Furthermore, the naive method
does not require CSI at the transmitter and is a robust method.} in
Fig. \ref{fig:low_SNR_worst}. Observe that the proposed method
outperforms the naive method at low SNR and the gain is contributed
by precoding.

\section{Conclusion}\label{sec:con} In this paper, we proposed a
robust lattice alignment design for $K$-user quasi-static MIMO
interference channels. We exploit the {\em structured interference}
and  propose a robust lattice alignment method for irrational
interference channels with imperfect CSI at all SNR regimes
as well as a two-stage decoding algorithm to decode the desired
signal from the structured interference space. For fairness, we
maximize the worst-case data rate by formulating the design of the
precoders, decorrelators, scaling coefficients and the interference
quantization coefficients as a mixed integer and continuous
optimization problem. By using an alternating optimization technique
and by incorporating imperfect CSI in the design, we derive
robust solutions with low complexity. Numerical results verify the
advantages of the proposed robust lattice alignment method compared
with various baseline systems.

\appendices \section{Proof of Lemma~\ref{lem:rate}}\label{app:rate}

From (\ref{eq:i_k^l}), the real part of the structured aggregate
interference is given by:
\begin{equation}
\begin{array}{ll}
&\Re\{\mathbf{I}_{k}^l\}=\left[\sum\nolimits_{i,n}\Re\{a_{i}^n\mathbf{x}_i^n\}\right]\text{mod
}\Lambda \\
=&\Big[\big(
\sum\nolimits_{i,n}\Re\{a_{i}^n\}\mathbf{t}_i^n-\Im\{a_{i}^n\}\mathbf{\widetilde{t}}_i^n\big)
\text{Mod } \Lambda -\\
&\quad
\sum\nolimits_{i,n}\Re\{a_{i}^n\}\mathbf{d}_i^n-\Im\{a_{i}^n\}\mathbf{\widetilde{d}}_i^n
\Big] \text{Mod } \Lambda.
\end{array}
\end{equation}
Thus, decoding $\Re\{\mathbf{I}_{k}^l\}$ is equivalent to decoding
$\mathbf{T}_k^l=\left[
\sum_{i,n}\left(\Re\{a_{i}^n\}\mathbf{t}_i^n-\Im\{a_{i}^n\}\mathbf{\widetilde{t}}_i^n\right)\right]
\text{Mod } \Lambda$. Note that the rate of $\mathbf{I}_{k}^l$ is
determined by the maximal rate of $\mathcal{L}_i^n,\forall a_i^n\neq
0$. Without loss of generality, the maximal data rate is denoted as
$R_{\max}$ and the corresponding nested lattice is given by
$\Lambda_{\max}$. The estimate of the $\mathbf{T}_k^l$ at receiver
$k$ is given by:
\begin{equation}
\begin{array}{ll}
\quad\quad\hat{\mathbf{T}}_k^l\\
=\left[
Q_{\Lambda_{\max}}\left(\Re\{\mathbf{y}_k^l\}+\sum\nolimits_{i,n}\Re\{a_{i}^n\}\mathbf{d}_i^n-\Im\{a_{i}^n\}\mathbf{\widetilde{d}}_i^n\right)\right]
\text{Mod }\Lambda\\
=\left[ Q_{\Lambda_{\max}}\left(\mathbf{T}_k^l+\mathbf{z}_k^l
\right)\right] \text{Mod }\Lambda,
\end{array}
\end{equation}
where
$\mathbf{z}_k^l=\Re\{\mathbf{u}_k^l\mathbf{Z}_k\}+\sum_{i,n}\Re\{(\mathbf{u}_k^l)^H\mathbf{H}_{ki}\mathbf{v}_i^n-a_i^n\}\Re\{\mathbf{x}_i^n\}-
\Im\{(\mathbf{u}_k^l)^H\mathbf{H}_{ki}\mathbf{v}_i^n-a_i^n\}\Im\{\mathbf{x}_i^n\}$.
From Lemma \ref{lem:erez-zamir}, $\Re\{\mathbf{x}_i^n\}$ and
$\Im\{\mathbf{x}_i^n\}$ are independently uniformly distributed over
$\mathcal{V}$ with
$\frac{1}{T}\mathbb{E}||\Re\{\mathbf{x}_i^n\}||^2=\frac{1}{T}\mathbb{E}||\Im\{\mathbf{x}_i^n\}||^2=P/2$,
and $\mathbb{E}[\mathbf{x}_i^n]=0$. From Lemma
\ref{lem:nazer-gastpar}, the density is upper bounded by the density
of an i.i.d. zero mean Gaussian vector $\mathbf{\widetilde{z}}_k^l$
whose variance $(\sigma_k^l)^2$ approaches
\begin{equation}
N_k^l =\frac{
||\mathbf{u}_k^l||^2}{2}+\frac{P}{2}\sum_{i,n}\Bigl|(\mathbf{u}_k^l)^H\mathbf{H}_{ki}\mathbf{v}_i^n-a_i^n\Bigr|^2.
\end{equation}

We set the volume of $\mathcal{V}_{\max}$ as
$\text{Vol}(\mathcal{V}_{\max})>\left(2\pi e (\sigma_k^l)^2
\right)^{T/2}$, since $\Lambda_k^l$ is AWGN good, the probability of
$\Pr\{\hat{\mathbf{T}}_k^l\neq\mathbf{T}_k^l\}$ goes to zero
exponentially in $T$
\cite{lattice:AWGN:2004,lattice:everywhere,CF:2009}. Let
$G(\Lambda)$ denote the normalized second moment of lattice
$\Lambda$, then
$\text{Vol}(\mathcal{V})=\left(\frac{\text{P}/2}{G(\Lambda)}
\right)^{T/2}$ \cite{lattice:everywhere}. From
(\ref{eq:lattice_rate}), the rate of the nested lattice code
$\mathcal{L}_{\max}$ is given by:
\begin{eqnarray}
R_{\max}=\frac{1}{T}\log\left(\frac{\text{Vol}(\mathcal{V})}{\text{Vol}(\mathcal{V}_{\max})}\right)
< \frac{1}{2}\log\left( \frac{P/2}{G(\Lambda)2\pi e (\sigma_k^l)^2}
\right).
\end{eqnarray}

Furthermore, for arbitrarily small $\xi>0$, and large enough $T$,
$G(\Lambda)2\pi e<(1+\xi)$\cite{lattice:everywhere}. By Lemma
\ref{lem:nazer-gastpar}, if $T$ is large enough,
$(\sigma_k^l)^2<(1+\xi)N_k^l$. As a result, the following rate is
achievable for $R_{\max}$:
\begin{equation}
\label{eq:xxx} \frac{1}{2}\log\Big(\frac{P}{||\mathbf{u}_k^l||^2+
P\sum_{i,n}|(\mathbf{u}_k^l)^H\mathbf{H}_{ki}\mathbf{v}_i^n-a_{i}^n|^2}\Big)-2\log(1+\xi).
\end{equation}
By choosing $\xi$ as sufficiently small, we can neglect the second
term, and the first term can be approached as desired. Similarly,
$\Im\{\mathbf{I}_{k}^l\}$ can also be successfully decoded from
$\Im\{\mathbf{y}_k^l\}$ with the same achievable data rate in
(\ref{eq:xxx}) for $R_{\max}$. Note that, if
$\mathbf{a}_k^l=\delta_{m}^d$ (all $a_i^n=0$ except that $a_m^d=1$),
$\mathbf{T}_k^l=\left[\mathbf{t}_m^d\right] \text{Mod } \Lambda =
\mathbf{t}_m^d$, and $\left[\mathbf{\widetilde{t}}_m^d\right]
\text{Mod } \Lambda = \mathbf{\widetilde{t}}_m^d$. Therefore,
 $\mathbf{x}_m^d$ can be decoded in stage I and completely nulled out at stage II.

Since the data rate $R_i^n$ is twice the lattice codes
$\mathcal{L}_i^n$ (due to the real and the imaginary parts), the
minimal achievable data rate of $R_i^n$ due to the effect of the CSI
error is given by
\begin{equation}
\label{eq:r_min}
\begin{array}{l}
R_{\min}=\\
\min_{\triangle_{ki}\in\mathcal{E}}\log\Big(\frac{P}{||\mathbf{u}_k^l||^2+
P\sum_{i,n}|(\mathbf{u}_k^l)^H(\hat{\mathbf{H}}_{ki}-\triangle_{ki})\mathbf{v}_i^n-a_{i}^n|^2}\Big).
\end{array}
\end{equation}

For given $\{\mathbf{a}_{k}^l,\mathbf{u}_k^l,\mathbf{v}_k^l\}$, a
lower bound of $R_{\min}$ is to choose $\triangle_{ki}$ maximizing
each term of the sum in the denominator, which is given by:
\begin{equation}
\begin{array}{ll}
&\max_{\triangle_{ki}\in\mathcal{E}}|(\mathbf{u}_k^l)^H(\hat{\mathbf{H}}_{ki}-\triangle_{ki})\mathbf{v}_i^n-a_{i}^n|^2\\
\Rightarrow&\max_{\triangle_{ki}\in\mathcal{E}}
|(\mathbf{u}_k^l)^H\triangle_{ki}\mathbf{v}_i^n|^2,
\end{array}
\end{equation}
where
\begin{equation}
\begin{array}{ll}
&|(\mathbf{u}_k^l)^H\triangle_{ki}\mathbf{v}_i^n|^2= \textrm{Tr}\bigl\{(\mathbf{u}_k^l)^H\triangle_{ki}\mathbf{v}_i^n(\mathbf{v}_i^n)^H\triangle_{ki}^H\mathbf{u}_k^l\bigr\}\\
=&\textrm{Tr}\bigl\{\mathbf{u}_k^l(\mathbf{u}_k^l)^H\triangle_{ki}\mathbf{v}_i^n(\mathbf{v}_i^n)^H\triangle_{ki}^H\bigr\}\\ \leq&\textrm{Tr}\bigl\{\mathbf{u}_k^l(\mathbf{u}_k^l)^H\bigr\} \textrm{Tr}\bigl\{\triangle_{ki}\mathbf{v}_i^n(\mathbf{v}_i^n)^H\triangle_{ki}^H\bigr\}\\
=&\textrm{Tr}\bigl\{\mathbf{u}_k^l(\mathbf{u}_k^l)^H\bigr\}\textrm{Tr} \bigl\{\mathbf{v}_i^n(\mathbf{v}_i^n)^H\triangle_{ki}^H\triangle_{ki}\bigr\}\\
\leq&\textrm{Tr}\bigl\{\mathbf{u}_k^l(\mathbf{u}_k^l)^H\bigr\}\textrm{Tr} \bigl\{\mathbf{v}_i^n(\mathbf{v}_i^n)^H\bigr\}\textrm{Tr}\bigl\{\triangle_{ki}^H\triangle_{ki}\bigr\}\\
=&||\mathbf{u}_k^l||^2||\mathbf{v}_i^n||^2\epsilon^2.
\end{array}
\end{equation}

Therefore, $R_{\min}$ given in (\ref{eq:r_min}) is lower bounded by:
\begin{equation}
\begin{array}{l}
R_{\min}\geq\mu_{k}^l=\\
\log\Big(\frac{P}{||\mathbf{u}_k^l||^2+
P\sum_{i,n}\bigl||(\mathbf{u}_k^l)^H\hat{\mathbf{H}}_{ki}\mathbf{v}_i^n-a_{i}^n|+
\epsilon||\mathbf{v}_i^n||\cdot||\mathbf{u}_k^l||\bigr|^2}\Big),
\end{array}
\end{equation}
which shows that $R_i^n<\mu_{k}^l$ is the sufficient condition for
the achievable data rate $R_i^n$ under imperfect CSI.

The proof of the Lemma \ref{lem:rate} is an extension of the results
in \cite{CF:2009}. However, there are the following differences.
Firstly, the results in \cite{CF:2009} are for perfect CSI while our
results are an extension of \cite{CF:2009} for imperfect CSI.
Secondly, the results in \cite{CF:2009} consider single stream relay
systems while our results are extended to multi-stream MIMO
interference channels. Finally, the results in \cite{CF:2009} are
analysis based whereas our solution is {\em optimization based}
where we shall optimize the precoders
$\mathbf{v}=\{\{\mathbf{v}_k^l\}_{k=1}^K\}_{l=1}^L$, first stage
decorrelators $\mathbf{u}=\{\{\mathbf{u}_{k}^l\}_{k=1}^K\}_{l=1}^L$,
the scaling coefficients $\mathbf{c}=\{\{c_k^l\}_{k=1}^K\}_{l=1}^L$
and the interference quantization coefficients
$\mathbf{a}=\{\{\mathbf{a}_k^l\}_{k=1}^K\}_{l=1}^L$. The detailed
formulation is illustrated in Section \ref{sec:sub_problem}.

\section{Convergence Proof of Subalgorithms A and B, and Top Algorithm 1}\label{app:alg}
We first provide the convergence proof of subalgorithm A. Note that
Given $\{\mathbf{v},\mathbf{a}\}$, after each iteration in
subalgorithm A (the alternating optimization between
$\mathbf{\widetilde{u}}_k^l$ and $c_k^l$), the data rate
$\widetilde{\mu}_k^l$ in (\ref{eq:R_k2}) is increasing.
Specifically, we have
$\widetilde{\mu}_k^l\left(\mathbf{\widetilde{u}}_k^l(m+1),c_k^l(m)\right)>\widetilde{\mu}_k^l\left(\mathbf{\widetilde{u}}_k^l(m),c_k^l(m)\right)$
in step 3 of subalgorithm A, and we have
$\widetilde{\mu}_k^l\left(\mathbf{\widetilde{u}}_k^l(m+1),c_k^l(m+1)\right)>\widetilde{\mu}_k^l\left(\mathbf{\widetilde{u}}_k^l(m+1),c_k^l(m)\right)$
in step 4 of subalgorithm B. Therefore,
\begin{equation}
\begin{array}{ll}
&\widetilde{\mu}_k^l\left(\mathbf{\widetilde{u}}_k^l(m+1),c_k^l(m+1)\right)\\
>&\widetilde{\mu}_k^l\left(\mathbf{\widetilde{u}}_k^l(m+1),c_k^l(m)\right)>\widetilde{\mu}_k^l\left(\mathbf{\widetilde{u}}_k^l(m),c_k^l(m)\right).
\end{array}
\end{equation}
As a result, subalgorithm A converges. In addition, we have
formulated subalgorithm B to a standard interior-point algorithm,
i.e., the barrier method \cite{Convex:2004}. Therefore the results
of subalgorithm B $\{\mathbf{v}^*,\mathbf{a}^*\}$ converge to a
$\varsigma$-optimal solution for fixed
$\{\mathbf{u},\mathbf{\widetilde{u}},\mathbf{c}\}$, and are subject
to the tolerance of $\varsigma$ \cite{Convex:2004}.

Note that
$R_{\min}=\min_{k,l}\big(\mu_k^l,\widetilde{\mu}_k^l\big)$, and by
using the above results we have
$\mu_k^l\big(\mathbf{u}_k^l(m+1),\mathbf{v}(m),\mathbf{a}(m)\big)>\mu_k^l\big(\mathbf{u}_k^l(m),\mathbf{v}(m),\mathbf{a}(m)\big)$,
and
$\widetilde{\mu}_k^l\big(\widetilde{\mathbf{u}}_k^l(m+1),c_k^l(m+1),\mathbf{v}(m),\mathbf{a}(m)\big)
>\widetilde{\mu}_k^l\big(\widetilde{\mathbf{u}}_k^l(m),c_k^l(m),\mathbf{v}(m),\mathbf{a}(m)\big)$
in step 2 of top algorithm 1, in other words,
$R_{\min}\big(\mathbf{u}(m+1),\widetilde{\mathbf{u}}(m+1),\mathbf{c}(m+1),\mathbf{v}(m),\mathbf{a}(m)\big)
>R_{\min}\big(\mathbf{u}(m),\widetilde{\mathbf{u}}(m),\mathbf{c}(m),\mathbf{v}(m),\mathbf{a}(m)\big)$.
Furthermore, we have
$t\big(\mathbf{u}(m+1),\widetilde{\mathbf{u}}(m+1),\mathbf{c}(m+1),\mathbf{v}(m+1),\mathbf{a}(m+1)\big)<
t\big(\mathbf{u}(m+1),\widetilde{\mathbf{u}}(m+1),\mathbf{c}(m+1),\mathbf{v}(m),\mathbf{a}(m)\big)$
in step 3 of top algorithm 1, where $t$ is given in
(\ref{eq:w_and_a_convex}). In other words,
$R_{\min}\big(\mathbf{u}(m+1),\widetilde{\mathbf{u}}(m+1),\mathbf{c}(m+1),\mathbf{v}(m+1),\mathbf{a}(m+1)\big)
>R_{\min}\big(\mathbf{u}(m+1),\widetilde{\mathbf{u}}(m+1),\mathbf{c}(m+1),\mathbf{v}(m),\mathbf{a}(m)\big)$.
Therefore we can conclude that
\begin{equation}\label{eq:iter}
\begin{array}{l}
\quad R_{\min}\big(\mathbf{u}(m+1),\widetilde{\mathbf{u}}(m+1),\mathbf{c}(m+1),\mathbf{v}(m+1),\mathbf{a}(m+1)\big)\\
>R_{\min}\big(\mathbf{u}(m+1),\widetilde{\mathbf{u}}(m+1),\mathbf{c}(m+1),\mathbf{v}(m),\mathbf{a}(m)\big)\\
>R_{\min}\big(\mathbf{u}(m),\widetilde{\mathbf{u}}(m),\mathbf{c}(m),\mathbf{v}(m),\mathbf{a}(m)\big).
\end{array}
\end{equation}

Since objective $R_{\min}$ increases after each iteration as
indicated in (\ref{eq:iter}), Algorithm 1 shall converge.

\section*{Acknowledgment}
The authors would like to thank F. Mckay at the Hong Kong University
of Science and Technology, and the anonymous reviewers for their
helpful comments and suggestions that significantly improved the
quality and the presentation of the paper.

\bibliographystyle{IEEEtran}
\bibliography{IEEEabrv,lattice_2nd}

\end{document}